\documentclass[aip,jcp,twocolumn]{revtex4}
\usepackage{blindtext}
\usepackage{graphicx}
\usepackage{subfig}   
\usepackage{float}    
\usepackage{amsmath}
\usepackage{enumerate}
\usepackage[sort&compress]{natbib}
\usepackage{amsmath}
\usepackage{amssymb}
\usepackage{setspace}

\usepackage{color}

\graphicspath{{./Figures/}}

\begin{document}
\title{Dissipative particle dynamics with energy conservation: isoenergetic integration and transport properties}
\author{Fatemeh A. Soleymani}
\author{Marisol Ripoll}
\email{m.ripoll@fz-juelich.de}
\author{Gerhard Gompper}
\author{Dmitry A. Fedosov}
\email{d.fedosov@fz-juelich.de}
\affiliation{Theoretical Soft Matter and Biophysics, \\Institute of Complex Systems and Institute for Advanced Simulation, \\
	Forschungszentrum J\"{u}lich, 52425 J\"{u}lich, Germany}
\date{\today}

\begin{abstract}
  Simulations of nano- to micro-meter scale fluidic systems under thermal gradients
  require consistent mesoscopic methods accounting for both
  hydrodynamic interactions and proper transport of
  energy. One such method is dissipative particle dynamics with
  energy conservation (DPDE), which has been used for various
  fluid systems with non-uniform temperature distributions. Despite the
  success of the method, existing integration algorithms 
  have shown to result in an undesired energy drift, putting into question
  whether the DPDE method properly captures properties of real
  fluids. We propose a modification of the velocity-Verlet
  algorithm with local energy conservation for each DPDE
  particle, such that the total energy is conserved up to machine
  precision. Furthermore, transport properties of a DPDE fluid are
  analyzed in detail.  In particular, an analytical approximation for
  the thermal conductivity coefficient is derived, which allows the selection
  of a specific value a priori. Finally, we provide approximate expressions for the dimensionless Prandtl
  and Schmidt numbers, which characterize fluid transport properties and can be adjusted independently by
  a proper selection of model parameters, and therefore, made
  comparable with those of real fluids. In conclusion, our results
  strengthen the DPDE method as a very robust approach for
  the investigation of mesoscopic systems with temperature inhomogeneities.
\end{abstract}
	
	
\maketitle
\section{Introduction}  

Simulations of mesoscopically structured systems, ranging from
supra-molecular assemblies and artificial self-propelled microswimmers
to the flow of biological cell suspensions in complex environments,
have become important in studies of a broad variety of biophysical,
biomedical, and engineering
applications~\cite{Pivkin_DPD_2011,Winkler_DRP_2014,Fedosov_MBF_2014,Elgeti_PMS_2015}.
This has driven a rapid development of mesoscopic simulation methods
employed to advance our understanding of such
systems~\cite{Schiller_MMS_2018,Gompper_APS_2009}.  Mesoscopic methods
usually neglect molecular details, but retain features of suspended
particles, such as deformability, inter-particle forces, or thermal
fluctuations.  Examples of mesoscopic methods are lattice Boltzmann
method (LBM)~\cite{McNamara_UBE_1988,He_LBM_1997,Succi_LBM_2001},
multiparticle collision dynamics
(MPC)~\cite{Malevanets_MSM_1999,Malevanets_SMD_2000,Gompper_APS_2009,Kapral_ACP_2008},
and dissipative particle dynamics
(DPD)~\cite{Hoogerbrugge_SMH_1992,Espanol_SMO_1995,Espanol_DPD_2017}.

Studies of many interesting phenomena require modeling of non-isothermal
environments, where temperature gradients are significant, and
transport of energy might be relevant. Examples include heat
transfer~\cite{Cleary_MCF_1998}, thermodiffusion in binary
mixtures~\cite{Sarman_HFM_1992,Wold_NMD_1999,Reith_NTD_2000,Artola_MTD_2008}, and colloidal
thermophoresis~\cite{Galliero_TMD_2008,Volpe_MSE_2011,Fedosov_EFC_2015,Luesebrink_TPC_2012,Tan_ATT_2017}.
Such problems generally rely on simulation methods which are able to represent
the system in a micro-canonical ensemble where energy is exactly
conserved. Various non-isothermal systems have been modeled using
molecular dynamics~\cite{Reith_NTD_2000,Sellan_EMD_2010}, thermal
LBM~\cite{McNamara_HDC_1997,Shan_SRB_1997},
MPC~\cite{Yang_TNS_2011,Yang_MTP_2014,Yang_TOM_2016}, and 
an energy-conserving version of the Monte Carlo method~\cite{Langenberg_eMC_2016}.  The original DPD
method~\cite{Hoogerbrugge_SMH_1992,Espanol_SMO_1995} is isothermal,
and has been extended to account for energy
conservation (DPDE)~\cite{Espanol_DPDE_1997,Avalos_DPDE_1997}. DPDE
has been successfully applied to simulate a number of thermal gradient
problems, such as natural~\cite{AbuNada_CHT_2010} and forced
convection \cite{Yamada_CHT_2011}, temperature-dependent fluid
properties~\cite{Li_EDPD_2014}, droplet flows~\cite{Yamada_DPD_2016}, and
thermophoretic Janus colloids~\cite{Fedosov_EFC_2015}.

Despite the success of DPDE, current formulation of the method still requires essential
improvements due to the difficulties in establishing an efficient and
precise integration algorithm. Several algorithms have been
employed in DPDE, including the Shardlow splitting scheme,
stochastic velocity-Verlet, and velocity-Verlet Shardlow splitting
algorithm~\cite{Avalos_DPDE_1999,Mackie_DPDE_1999,Lisal_DPD_2011,Larentzos_DPDE_2014,Li_EDPD_2014,Homman_DPDE_2016}.
Even though these algorithms have been developed as
energy-conserving methods, a general problem of all the integration
schemes is that they yield a significant net energy drift. 
This drift can be reduced to a large extent by
decreasing the integration timestep. This makes the method less
attractive, as it becomes more CPU-time consuming and some remaining drift does not completely
disappear, rendering long-time integration difficult. Furthermore, it is
not fully clear whether the DPDE method can faithfully represent
important features (e.g. thermal diffusivity, momentum transport) of
real fluids, as the transport properties of a DPDE fluid are not
directly predetermined.

In this paper, we suggest a modification of the well-known velocity-Verlet algorithm, which results in energy conservation up to the order 
of machine precision,  and completely eliminates energy drift. Within this algorithm, changes in kinetic and potential energies are 
counterbalanced by the internal energy at the level of each DPDE particle, such that the energy remains exactly conserved. 
We investigate the performance of the algorithm for several choices of conservative interactions between particles and two 
different formulations of the inter-particle heat conduction. The different heat-conduction terms exhibit nearly identical performance, 
while different choices for conservative interactions affect the fluid compressibility. Furthermore, an analytical approximation 
of the overall thermal conductivity $\kappa$ of a DPDE fluid is derived, which allows its direct estimation for selected simulation
parameters. The thermal conductivity is governed by two contributions: (i)~direct heat conduction and (ii)~diffusive heat transport, whose relative ratio 
can be estimated from the analytical approximation of $\kappa$. Finally, we study the dimensionless Prandtl and Schmidt 
numbers characterizing fluid transport properties, and show that they can be independently adjusted to match those 
of real fluids.            

The paper is organized as follows. Section~\ref{models and methods} provides details about the DPDE method and the employed 
integration algorithm, with the details given in Appendix~\ref{sec:alg}. In section~\ref{sec:results}, the DPDE method is validated and its performance 
is studied under thermal gradient conditions. In section~\ref{sec:results}, we also derive the analytical approximation of the fluid thermal conductivity 
coefficient and discuss transport properties of simulated liquids in comparison to real fluids. 

\section{Method}
\label{models and methods}

\subsection{Mass and momentum evolution}
\label{internalTemp}

In the same spirit as in the isothermal DPD
method~\cite{Hoogerbrugge_SMH_1992,Espanol_SMO_1995,Marsh_FPB_1997}, DPDE particles
are mesoscopic entities, which represent small fluid volumes
containing numerous atoms or molecules. The $N$
constituent particles are characterized by their positions
$\textbf{r}_{i}$ and velocities $\textbf{v}_{i}$. Furthermore, 
each of the DPDE particles is characterized by an
additional variable accounting for the internal energy $\epsilon_{i}$
of the small fluid volume.  This energy is connected to an
internal temperature $T(\epsilon)$ via an entropy function
$S(\epsilon)$ as $\partial S(\epsilon)/\partial\epsilon =
1/T(\epsilon)$.  A simple choice for the entropy function is that of
an ideal solid, i.e. $S(\epsilon) = c_{v}\ln
(\epsilon)+\mathrm{const}$, where $c_{v}$ is the heat capacity at
constant volume. This choice results in a straightforward linear relation between internal
energy and temperature as
$\epsilon_{i}=c_{v}T_{i}$~\cite{Espanol_DPDE_1997,Avalos_DPDE_1997,Ripoll_DPDE_1998,Qiao_SHC_2007}.

Time evolution of particle characteristics (i.e. position, velocity, and internal energy) is
governed by the Newton's equation of motion and heat equation  
\begin{equation}\label{equation_motion}
\frac{\textrm{d}\textbf{r}_{i}}{\textrm{d}t}=\textbf{v}_i, 
\quad \frac{\textrm{d}\textbf{v}_{i}}{\textrm{d}t}=\frac{1}{m_i}\textbf{F}_{i},
\quad  c_{v} \frac{\textrm{d} {T}_{i}}{\textrm{d}t} = q_{i}. 
\end{equation}
Here,
$m_i$ is the particle mass, $\textbf{F}_{i}$ is the total force, and
$q_{i}$ is the total heat rate. The force $\textbf{F}_{i}$ on particle $i$ is a sum of three pairwise
interactions with neighboring particles $j$ given by
\begin{equation}\label{total_force}
\textbf{F}_{i}=\sum\limits_{j\ne i}\left(\textbf{F}^{C}_{ij}+\textbf{F}^{D}_{ij}+\textbf{F}^{R}_{ij}\right).
\end{equation}
The three contributions correspond to 
the conservative, dissipative, and random forces which take the form
\begin{equation}\label{eq:force}
\begin{aligned}
		 &\textbf{F}_{ij}^{C} = a_{ij} \omega ^{C}(r_{ij}) \hat{\textbf{r}}_{ij}, \\ 
		& \textbf{F}_{ij}^{D} = -\gamma_{ij} \omega ^{D}(r_{ij}) (\textbf{v}_{ij}\cdot\hat{\textbf{r}}_{ij}) \hat{\textbf{r}}_{ij},   \\
		 &\textbf{F}_{ij}^{R} = \sigma_{ij} \omega ^{R}(r_{ij}) \xi _{ij} \Delta t^{-1/2}\hat{\textbf{r}}_{ij}. 
\end{aligned}
\end{equation}
Here, $\textbf{r}_{ij} = \textbf{r}_{i} - \textbf{r}_{j}$,
$\textbf{v}_{ij} = \textbf{v}_{i} - \textbf{v}_{j}$, $r_{ij} = |\textbf{r}_{ij}|$,
$\hat{\textbf{r}}_{ij} = \textbf{r}_{ij} / r_{ij}$, and $\Delta t$ is
the integration timestep.
The conservative-force coefficient $a_{ij}$ controls fluid
compressibility. The coefficients $\gamma _{ij}$ and $ \sigma _{ij}$
represent friction and noise amplitudes that are connected to each
other through the fluctuation-dissipation
relation~\cite{Espanol_DPDE_1997,Avalos_DPDE_1997,Marsh_DBH_1998} as
\begin{equation}\label{gamma_sig}
\gamma _{ij} = \dfrac{\sigma ^{2}_{ij}}{4k_{B}}\left( \dfrac{1}{T_{i}}+\dfrac{1}{T_{j}}\right)
\end{equation}
with the Boltzmann constant $k_{B}$. The random force is
determined by $\xi _{ij}$, a symmetric Gaussian random variable
(i.e. $\xi _{ij} = \xi _{ji}$) with zero mean and unit variance.
Generally, $\sigma_{ij}$ is selected as a constant
and $\gamma_{ij}$ is calculated according to Eq.~(\ref{gamma_sig}).
The interaction strengths in Eq.~(\ref{eq:force}) are further
controlled by the weight functions which are most commonly chosen as
\begin{equation}
  \omega(r_{ij}) = 
  \begin{cases}
    \left(1-\dfrac{r_{ij}}{r_{c}}\right)^{2s} & r_{ij} \leq r_{c},\\
    0 & r_{ij} > r_{c},
  \end{cases} 
\end{equation}
where $r_{c}$ is the cutoff radius and $s>0$ is an exponent controlling
the interaction strength.  While the dissipative and the random
functions are linked via the fluctuation dissipation theorem as
$\omega^{D}= \big(\omega^{R}\big)^{2}= \omega$
\cite{Espanol_SMO_1995}, the choice of the conservative function
$\omega^{C}$ is independent.

\subsection{Energy evolution: isoenergetic integration}
\label{sec:en_eq}
Time evolution of the internal temperature (or the corresponding
internal energy) is governed by the third equality in
Eq.~(\ref{equation_motion}).  
The amount of heat that each particle 
receives (or loses) from its close neighbors per unit time can be expressed 
as 
\begin{equation} \label{eq:heat}
q_{i}=\sum\limits_{j\ne i} \left( q_{ij}^{HC} + q_{ij}^{VH} \right),
\end{equation} 
where $q_{ij}^{HC}$ corresponds to {\em heat conduction}, and
$q_{ij}^{VH}$ to {\em viscous heating}.  

DPDE particles are characterized by an intrinsic temperature, such that 
every pair of particles exchanges some heat by
conduction, proportional to the interparticle
temperature difference. 
Originally~\cite{Espanol_DPDE_1997,Avalos_DPDE_1997}, this
term has been proposed to be proportional the difference of the inverse
temperatures,
\begin{equation}\label{q_E}
	q^{A}_{ij} = \kappa^{A}_{ij} \omega^{H}(r_{ij}) 
\left( \frac{1}{T_{i}}-\frac{1}{T_{j}} \right) 
+ \alpha^{A}_{ij} \zeta_{ij} \left[\omega^{H}(r_{ij})/\Delta t\right]^{1/2}, 
\end{equation}
where $\kappa^{A}_{ij}$ and $\alpha^{A}_{ij} $ are heat-conduction coefficients of 
the deterministic and random terms, respectively. 
To guarantee energy conservation, each term in Eq.~(\ref{q_E}) needs to be
antisymmetric under particle interchange. The deterministic term is
antisymmetric by construction, and the fluctuating factor $\zeta _{ij}$
is defined as an antisymmetric Gaussian variable with zero mean
and unit variance, such that $\zeta_{ij} = -\zeta_{ji}$. Note that $\zeta_{ij}$ is
completely uncorrelated with $\xi_{ij}$. Detailed balance imposes the
relation between the deterministic and fluctuating terms. Thus, the
heat conduction coefficients are connected by the fluctuation-dissipation relation
$\left(\alpha^{A}_{ij} \right)^{2} = 2k_{B}\kappa^{A}_{ij}$, where 
$\kappa^{A}_{ij} = \kappa^{A}_{0} c_{v}^{2}(T_{i}+T_{j})^{2}/4k_{B}$
with the constant nominal strength $\kappa^{A}_{0}$.  
The weight function $\omega^H$
can generally be chosen independent of the other weight functions,
although we select $\omega^{H}=\omega^{D}$ here for simplicity. 

An alternative form for $q_{ij}^{HC}$ was proposed later~\cite{Ripoll_MSF_2005,Ripoll_PLT_2005}, 
where the heat conduction is directly proportional to the 
temperature difference between particles as expected from the Fourier law, 
\begin{equation}\label{q_RE}
  q^{B}_{ij} = \kappa^{B}_{ij} \omega^{H}(r_{ij}) (T_{j} - T_{i}) 
 + \alpha^{B}_{ij}\zeta_{ij}  \left[\omega^{H}(r_{ij})/ \Delta t\right]^{1/2}. 
\end{equation}
Here,  the fluctuation-dissipation relation requires that
$\left(\alpha^{B}_{ij}\right)^2 = \kappa^{B}_{ij} T_{i}T_{j}$ with
$\kappa^{B}_{ij} = c_{v} \kappa^{B}_{0}$. 
The correspondence between these two models for $q_{ij}^{HC}$ can be
achieved for $\kappa^{B} _{0} = \kappa^{A}_{0}c_{v}/ k_{B}$, if the
local temperature differences are not large, i.e. $T_i \approx T_j$. 
Quantitative comparison between these two models has not been
reported yet and will be presented later.

The second contribution to the heat transport in Eq.~(\ref{eq:heat})
is the viscous heating term $q_{i}^{VH}=\sum_{i \neq j}q_{ij}^{VH}$,
which represents the work done by conservative and dissipative forces. Thus, it accounts
for the variation in mechanical energy as $q_{i}^{VH}=\delta
{E_{mec,_i}} = \delta P_i + \delta K_i$, where $P_{i}$ and $K_i$ are
the potential and kinetic energies whose changes can be calculated as 
\begin{eqnarray}
\delta P_i &=& - \sum_{i \neq j}\textbf{F}^{C}_{ij}\cdot\delta \textbf{r}_{ij} \\
\delta K_i &=&  m \sum_{i} \textbf{v}_{i} \cdot\text{d}\textbf{v}_{i} +
m \sum_{i} \text{d}\textbf{v}_{i}\cdot\text{d}\textbf{v}_{i}. \nonumber
\end{eqnarray}
The second sum in the variation of kinetic energy is a consequence of
the stochastic contribution. By considering the equation of
motion~(\ref{equation_motion}) and the form of the DPD forces in
Eq.~(\ref{eq:force}), the change in mechanical energy can be expressed
explicitly through the DPD force
parameters~\cite{Espanol_DPDE_1997,Avalos_DPDE_1997}. In fact, such
non-trivial expressions for $q_{i}^{VH}$ are often integrated directly
using various existing methods such as Shardlow or
Trotter schemes~\cite{Shardlow_DPD_2003,De_Fabritiis_ENI_2006,Serrano_STA_2006}.
However, we do not use such an expression, but calculate
$q_{i}^{VH}$ by a direct tracking of changes in the kinetic and potential
energies for every particle at each timestep (see
Appendix~\ref{sec:alg} for details about the integration algorithm).
In this way, viscous heating is accounted for and the energy is
conserved locally as well as globally.  In a study ~\cite{Lisal_DPD_2011} focused on the
performance of different variants of the Shardlow splitting
scheme, a related idea has already been
outlined, but the energy conservation has been implemented at the
level of particle pairs in contrast to the particle level proposed
here, and the conservation of potential energy was not considered. A
further important advantage of our integration approach is that it
allows much easier parallelization than more sophisticated
algorithms, such as Shardlow or
Trotter schemes~\cite{Larentzos_DPDE_2014,Homman_DPDE_2016}, making it a
powerful candidate for applications.

\subsection{Simulation setup and parameters}
\label{sec:pmt}

Simulation units are selected to be the particle mass $m=1$, the
length of the cutoff radius $r_c=1$, and the unit of energy $k_BT_0=1$
with the reference temperature $T_0$.  In this way, $\tau = r_{c}
\sqrt{m/(k_{B}T_{0})} = 1$ corresponds to the unit of time.  The
default average temperature and number density of fluid particles are
set to $\bar{T}=T_0$ and $\bar{\rho} = 3/r_{c}^3$, respectively,
unless specified otherwise. 
The DPDE forces employ $\omega^{C}=\omega$ with $s=0.5$ as used 
commonly in DPD, $\sigma = 3k_{B}T_{0}\sqrt{\tau}/r_{c}$, and
$\omega^{D}=\omega^{H}=\omega$ with $s=0.25$ whose value has been shown to
increase the fluid viscosity in comparison to a frequently used value of
$s=1$~\cite{Fan_DNA_2006,Fedosov_VL_2008}. 
The specific heat is set to $c_v = 200k_B$ and the nominal strengths
of interparticle heat conductivities are $\kappa^{A}_{0} = 0.001/\tau$
and $\kappa^{B}_{0} = 0.2/\tau$ for the two heat-rate models presented
above.  The dimensionless heat capacity $\beta=c_{v}/k_{B}$
characterizes the number of internal degrees of freedom of a DPDE
particle, which in general should be large
enough~\cite{Ripoll_MSF_2005}.

The conservative-force coefficient $a_{ij}$ is typically taken as a
constant in DPDE, following a similar choice in the classical DPD
method. However, a dependence of $a_{ij}$ on
temperature has been suggested to represent better temperature-dependent fluid
properties~\cite{Li_EDPD_2014,Groot_DPD_1997}. We consider 
three cases: (i) the ideal gas equation of state case with
$a^{C}_{ij}=0$, (ii) the case with a constant coefficient
$a^{C}_{ij}=a_{0}$, and (iii) the case with a temperature-dependent
coefficient $a^{T}_{ij} = a_{0}(T_{i}+T_{j})/2\bar{T}$, for which the
default constant parameter is taken to be $a_{0} =
15k_{B}T_{0}/r_{c}$. Different combinations of the heat conduction 
term (i.e. $q^{A}_{ij}$ and $q^{B}_{ij}$) and the conservative-force 
coefficient (i.e. $a^{C}_{ij}$ and $a^{T}_{ij}$) employed in simulations 
will be denoted as e.g. $a^C q^A$. Note that the default combination 
is $a^T q^A$ if not stated otherwise. A default timestep
corresponds to $\Delta t = 0.005\tau$. Simulations are performed with
periodic boundary conditions in a domain with dimensions
$L_{x}=20r_{c}$, $L_{y}=L_{z}=10r_{c}$.  All other parameters will be
specified in the text whenever necessary.

\section{Results} 
\label{sec:results}

\subsection{Homogeneous temperature}
\label{valid}

\paragraph{Energy conservation.-}
One of the main issues of the DPD method is the establishment of an efficient and consistent 
algorithm for the integration of the DPD equations, which has been, and still is 
treated in a large number of related
studies~\cite{Groot_DPD_1997,Nikunen_DPD_2003,Thalmanna_TDA_2007,Litvinov_DPD_2010,Leimkuhler_DPD_2015,Yamada_DPD_2018}.
Although DPDE was proposed as an energy-conserving method, so far all investigations,
in which the viscous heating term $q_{ij}^{VH}$ is calculated explicitly through the system
parameters~\cite{Shardlow_DPD_2003,De_Fabritiis_ENI_2006,Espanol_ECG_2016}, 
have reported different deviations from energy-conserving behavior for
various timesteps and integration schemes. 
For example, Homman {\it et al.}~\cite{Homman_DPDE_2016} report an
energy drift within a range $1-2 \times 10^{-5} k_{B}T_0$
for $\Delta t = 0.006 \tau$ and a total integration time of $10\tau$,
while L{\'i}sal {\it et al.}~\cite{Lisal_DPD_2011,Larentzos_DPDE_2014} find 
an energy drift on the order of $10^{-4}$,
even though the viscous heat rate was not  explicitly calculated. Furthermore, 
Li {\it et al.}~\cite{Li_EDPD_2014} demonstrate
fluctuations in the total energy on the order of $ 10^{-3}
k_{B}T_0$, even though no significant energy drift within the
total integration time was detected.

As explained in Sec.~\ref{sec:en_eq}, the integration algorithm adopted here is
fundamentally different from those used previously, since the variation of
energy per particle is directly monitored and used for the calculation of the
viscous-heating term, avoiding its approximation in terms of
parameters such as the relative velocities or positions of all
neighboring particles. 
We have verified the conservation of total energy for a range of
timesteps $\Delta t / \tau \in [0.005; 0.05]$ and for two system
sizes, the one with default values and another one approximately $10$
times larger.  Maximum error and fluctuations in the total energy are
found to be on the order of $10^{-14}k_{B}T_0$, which demonstrates that
our DPDE algorithm conserves the total energy by construction (see
Appendix \ref{sec:alg}), and therefore, the error in total energy is
directly associated with the machine precision.

Note that in order to evaluate the total energy, the
mechanical energy needs to be considered together with the internal
energy $I_E$.  The internal energy constitutes the dominant
contribution to the total energy, since $I_E\approx N c_v T_0$ and
$c_v \gg 1$, and the kinetic energy is $K\simeq 3N k_{B}T_0/2$
(i.e. $I_E / K \simeq 2c_v/3k_B$), while our simulations show that the
potential energy is $P\simeq 2K$ for the default conservative
interactions.

\begin{figure}[h]
    \includegraphics[width=0.48\textwidth]{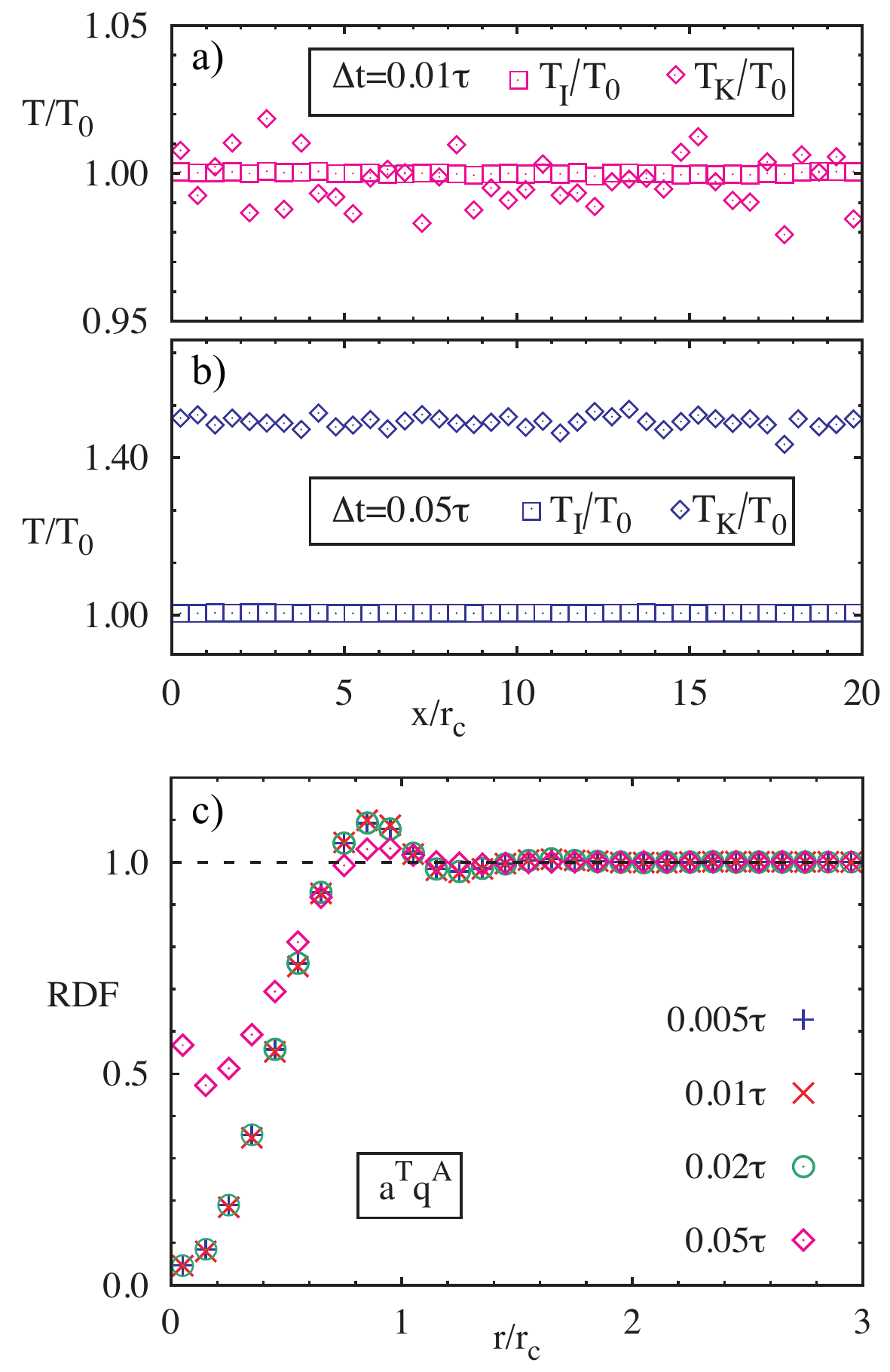}
    \caption{a,b)~Comparison of internal $T_{I}/T_0$ and kinetic $T_{K}/T_0$
      temperatures for two different time steps as a function of position 
      within the simulation domain. 
      c)~Radial distribution function (RDF) of fluid particles for
      various timesteps, showing that for $\Delta t \lesssim 0.02\tau$
      the RDF is independent of the timestep.
      \label{fig:temp_rdf}}
\end{figure}

\paragraph{Temperature definition.-}	
In the DPDE method, three temperatures can be defined which, for
consistency, should be equivalent. The internal temperature $T_I$ is
 computed by averaging over the internal particle temperatures
$\langle T_{i} \rangle$; the kinetic temperature $T_K$ is calculated
by averaging the kinetic energy of all particles as $\langle m_i
v_i^{2}/3 \rangle$, and both can also be compared to the
reference temperature $T_0$.  Comparison of the simulation results is
shown in Fig.~\ref{fig:temp_rdf}a,b for two different timestep
values. $\Delta t = 0.01\tau$ results in consistent internal
and kinetic temperatures, while $\Delta t = 0.05\tau$ leads to
values of $T_{I}$ and $T_K$ which differ by nearly 50\%, indicating that this timestep is
too large.

Figure~\ref{fig:temp_rdf}a also shows that the variance of kinetic
temperature is markedly higher than that of the internal temperature, 
which can be rationalized by differences in temperature distributions. 
The variance of $T_K$ is determined by the
Maxwell-Boltzmann distribution of particle velocities and by the
number of particles used for $T_K$
averaging~\cite{Lopez_DMD_2007}. The variance of $T_I$ is given by
$\sigma^{2}_{T_{I}} = \sigma^{2}_{\epsilon} / c_{v}^{2}$, where
$\sigma^{2}_{\epsilon}$ is the variance of internal energy, which can
be estimated from the distribution of $\epsilon$ in the micro-canonical
ensemble~\cite{Ripoll_thesis_2002}.

\paragraph{Radial distribution function.-}	
The measurement of the radial distribution (RDF) function is another 
test to elucidate the appropriateness of the
employed parameters, in particular the timestep.
Figure~\ref{fig:temp_rdf}c demonstrates that $\Delta t \lesssim
0.02\tau$ leads to timestep-independent RDF within the fluid.
Even though Fig.~\ref{fig:temp_rdf} presents results only for the
$a^T q^{A}$ model, other combinations of the conservative-force 
coefficient and the heat conduction term have also been
tested, resulting in a similar conclusion.  Note that the recommended
values of the timestep are about one order of magnitude larger than
those employed with previous integration methods~\cite{Li_EDPD_2014}.

\subsection{Temperature gradients}
\label{thermalG}

Different strategies have been employed to simulate thermal
gradients~\cite{Sarman_HFM_1992,Reith_NTD_2000,Lusebrink_TIH_2012},
which can be used in DPDE as well. One of these methods, the
velocity exchange algorithm~\cite{Reith_NTD_2000}, exactly
conserves the overall energy by interchanging the velocity of the
fastest particle within a pre-determined cold slab with the slowest
particle of a hot slab. Alternatively, the average temperature of
two pre-defined layers (or regions) can be set to two unequal
values~\cite{Pooley_SRD_2005,Yang_TPF_2013}.  Furthermore, 
simulations with periodic boundary conditions and temperature 
gradients often employ hot and cold slabs, in which 
heat is respectively injected and removed, 
mimicking the contact with reservoirs at different temperatures. In
our simulations, heat is injected into the hot slab ($10<x/r_c<10.5$)
at a constant heat rate $2J_0 A$, where $J_0$ is the heat flux and
$A=L_yL_z$ is the cross-sectional area of the domain, while heat is
removed from the cold slab ($0<x/r_c<0.5$) at the same rate. Thus, a
fixed amount of heat $Q=2J_0 A \Delta t$ is uniformly added or removed
every timestep from the internal energy of all particles in the volume
of a given slab. The factor $2$ is due to periodic
boundary conditions such that the injected heat $Q$ travels to the
left and to the right from the hot slab. Hence, every particle in the
hot (cold) slab receives (losses) a heat of $Q/N_h$ ($Q/N_c$) per timestep, where $N_h$ ($N_c$) is a
time-varying number of particles in the hot (cold) slab. 
In this way, the total energy in the simulation domain is conserved and the temperature
gradient can be indirectly regulated by changing the heat rate or
flux.

\begin{figure}[h]
  \begin{center}
    \includegraphics[width=0.48\textwidth]{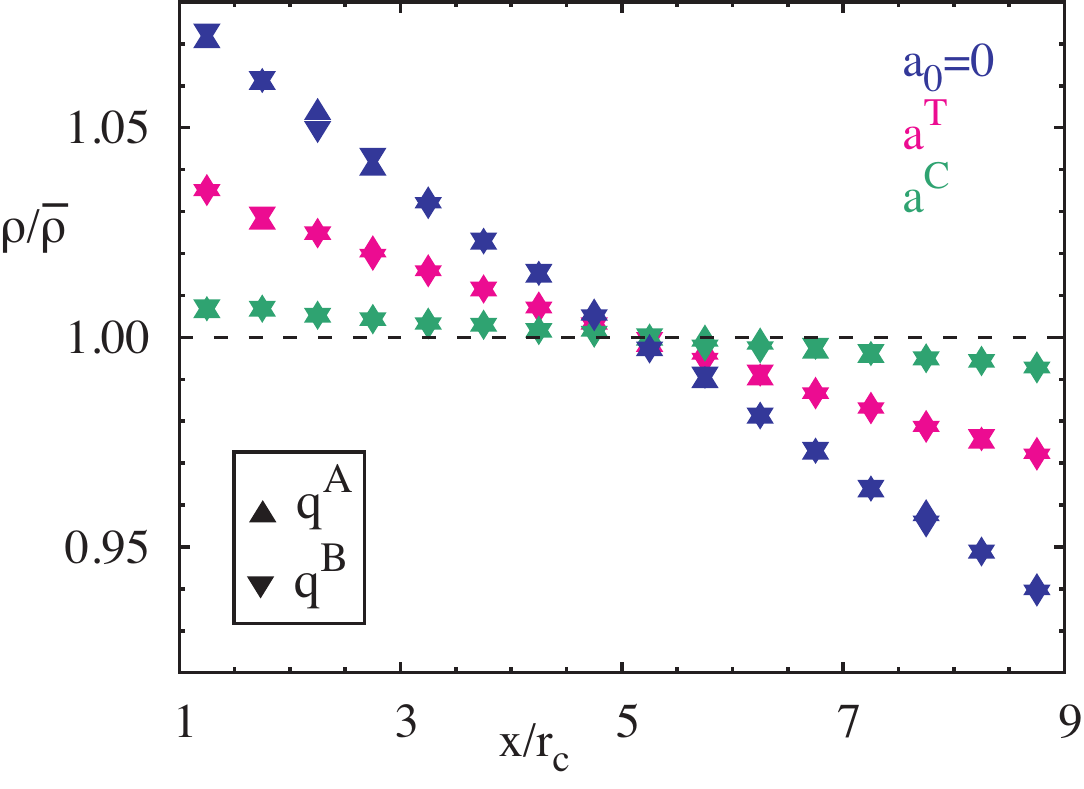} 
    \caption{Steady-state density profiles for six
      different models with the same imposed thermal
      gradient. Two heat conduction models $q^A$ in
      Eq.~(\ref{q_E}) and $q^B$ in Eq.~(\ref{q_RE}) are
      combined with three choices of the conservative
      force as introduced in Sec.~\ref{sec:pmt}.}
    \label{fig:temp_dens} 
	\end{center}
\end{figure}

\paragraph{Equation of state.-}
First, we test the development of temperature gradient for the two
models of heat conduction term, including $q^A$ in Eq.~(\ref{q_E})
and $q^B$ in Eq.~(\ref{q_RE}), together with three different choices
of conservative interactions, using $a^T$ and $a^C$ models with
$a_{0} = 15k_{B}T_{0}/r_{c}$, and a model with $a_{0} = 0$. In the case
$a_{0} = 0$, the fluid has an ideal-gas equation of state with $p=\rho
k_B T$, where $p$ is the pressure and $\rho$ is the density of the
fluid. For comparison, isothermal DPD with the $a^C$ model yields an
equation of state where pressure varies quadratically in fluid density~\cite{Groot_DPD_1997}.  
The six possible model combinations are tested for the same linear
temperature profile, which is acquired by properly adjusting $J_0$. 
Thus, the profile $T(x)/T_0=0.018x/r_{c}+0.91$ with
$T_{min}/T_0\approx 0.93$ and $T_{max}/T_0\approx 1.073$ is obtained
for the models $a^T$, $a^C$, and $a_{0} = 0$, with $J_0\tau r_c^2/(k_BT_0)=2.7$,
$2.75$, and $2.8$, respectively. 
Note that the temperature profile is independent of the choice of
the heat conduction term.  Figure~\ref{fig:temp_dens} presents
steady-state density profiles for the different models with the same
temperature profile, which are nearly linear for all cases. The
strongest gradient in fluid density is found for the $a_{0} = 0$
model, while the $a^C$ model leads to the least variation in $\rho$.
This means that the $a^C$ model yields the least compressible fluid,
while the $a^T$ and $a_{0} = 0$ models result in more compressible
fluids. Note that there is no noticeable difference between the $q^A$
and $q^B$ models for the heat conduction term in DPDE.

\begin{figure}[h!]
  \includegraphics[width=0.9\columnwidth]{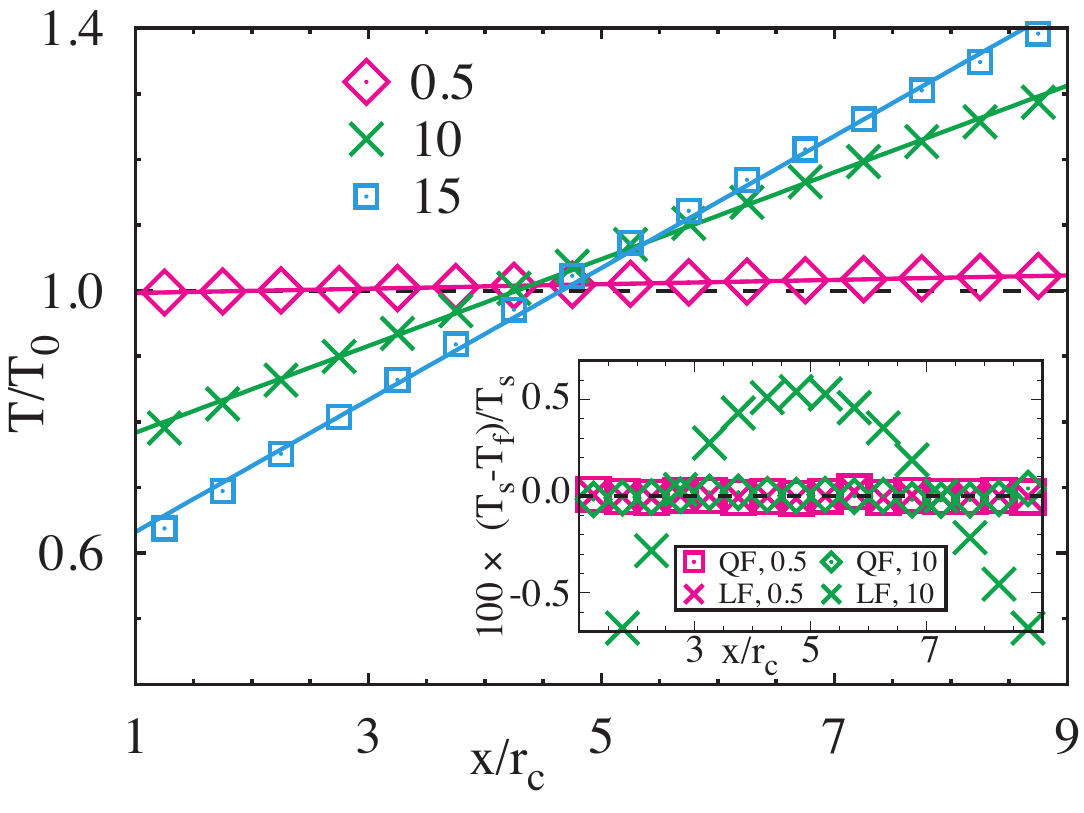}
  \caption{Temperature profiles 
    for three different $J_0\tau r_c^2/(k_BT_0)$
    values. Inset shows deviations of the temperature profile from linear (LF)
    and quadratic (QF) fits in percents for two different heat flux values.}
\label{fig:cond}
\end{figure}

\paragraph{Heat conductivity.-}
The Fourier law of heat conduction in one dimension, 
\begin{equation} \label{eq:fourier} J =
  \dfrac{1}{A}\dfrac{\mathrm{d}Q}{\mathrm{d}t} = \kappa
  \dfrac{\mathrm{d}T}{\mathrm{d}x},
\end{equation}
allows the determination of the thermal
conductivity $\kappa$ in simulations with fixed $J_0$ by analyzing
the resulting temperature gradient. Figure~\ref{fig:cond} shows three
temperature profiles for different $J_0$. All temperature curves
exhibit a nearly linear dependence; however, for the highest $J_0$
value, the simulation data slightly deviate from
the corresponding linear fit. The deviation of simulated temperature
from a linear fit is quantified in the inset of Fig.~\ref{fig:cond}. The
temperature profile for $J_0\tau r_c^2/(k_BT_0)=10$ can be fitted well
by a quadratic function.  Nevertheless, the deviation of temperature
from the linear fit is small and remains within $~2\%$. A large enough
heat flux clearly leads to a strong temperature gradient with local
changes in fluid density and structure characterized by RDF. This
in turn may affect local heat conductivity, as it depends on
inter-particle distances, resulting in a slightly non-linear
temperature profile.

\begin{figure}[h]
	\includegraphics[width=0.9\columnwidth]{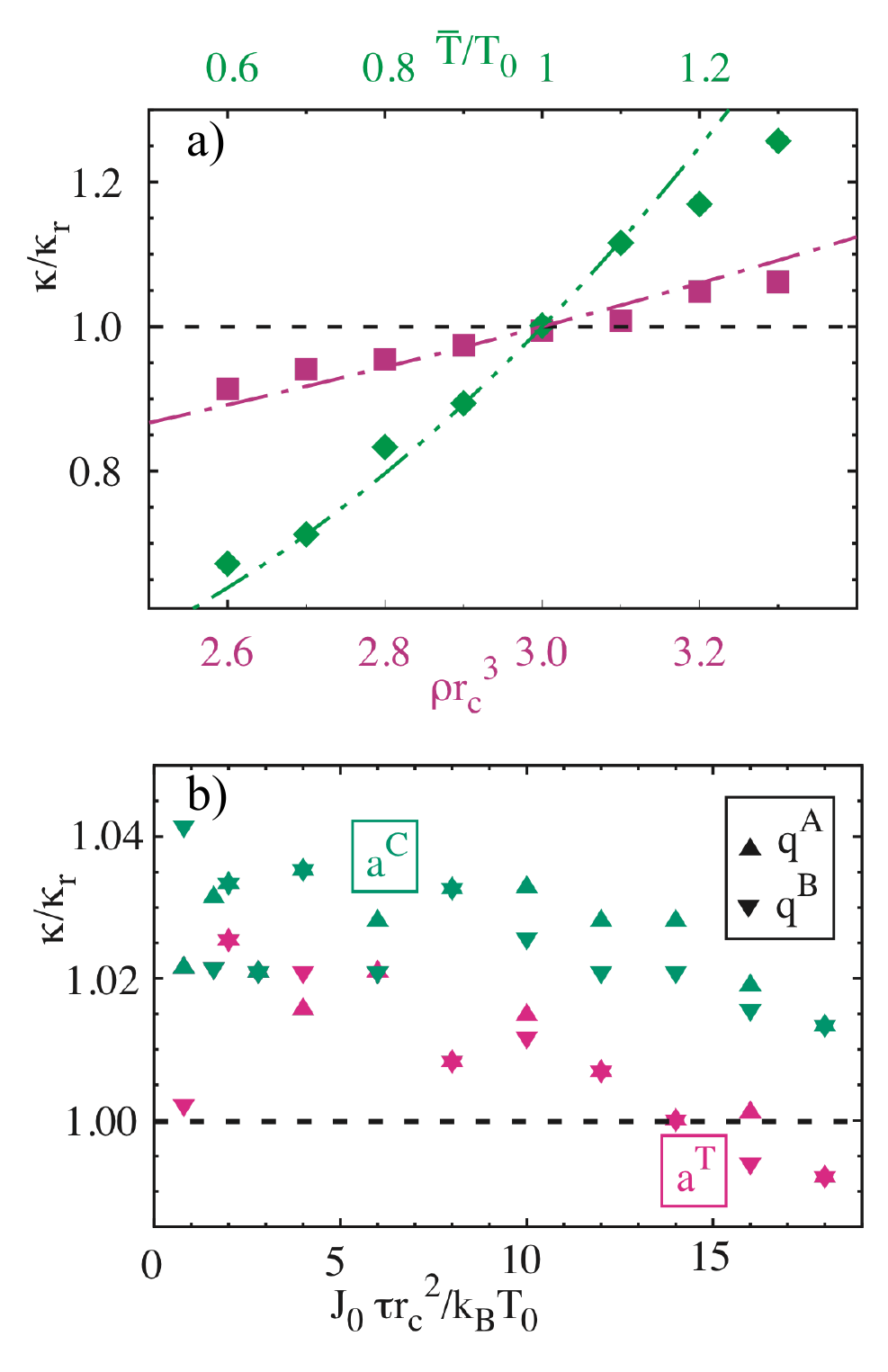}
  \caption{a)~Normalized heat conduction coefficient $\kappa/\kappa_r$
    as a function of temperature and density for relatively weak
    thermal gradients ($\Delta T/T_0\lesssim 0.05$). The normalizing
    factor $\kappa_r=146k_B/(\tau r_c)$ is calculated for the
    reference values $\bar{\rho}=3/r_c^3$ and $\bar{T}=T_0$. Dashed
    dotted lines and the value of $\kappa_r$ are obtained from the
    analytical prediction in Eq.~(\ref{con-equ}). b)~Normalized heat
    conduction coefficient as a function of the input heat flux for
    different models.}
\label{fig:kappa}
\end{figure}

Figure~\ref{fig:kappa}a presents the dependence of heat conductivity
$\kappa$ on the average fluid temperature and density for
the reference parameters, using a small value of $J_0\tau r_c^2/(k_B
T_{0})=0.25$.  Thermal conductivity increases when average fluid density
or temperature increase, which is qualitatively consistent with other
experimental and theoretical
investigations~\cite{Chapman_NUG_1990,Guildner_TCG_1975,Hanley_VTC_1974,Ziebland_TCL_1955}.
We will discuss these dependencies in more detail in
Sec.~\ref{theory}, where an analytical expression for $\kappa$ is
derived. Figure~\ref{fig:kappa}b shows the dependence of thermal
conductivity $\kappa$ on heat flux $J_0$ for different models of
the heat conduction term and conservative interactions. Since $\kappa$
depends on local temperature and density, it is obtained here by
fitting temperature profiles within a small region where particle
densities and temperatures are very close to their average
values. Differences in $\kappa$ values for the $q^A$ and $q^B$ conductivity terms
are not really significant, while a temperature-dependent conservative-force 
coefficient $a^T$ results in slightly smaller heat conductivities in
comparison to the constant conservative force with $a^C$. Heat conductivity
exhibits a slight decrease with increasing input heat flux
$J_0$. However, this decrease is within $3-4\%$, indicating that
$\kappa$ in the DPDE method is rather robust and nearly independent of
the heat flux.

The fact that the DPDE model has a finite value of heat conductivity
implies that only a limited amount of heat can be transferred,
which determines a maximum heat flux $J_{max}$.  This means that if lager heat
fluxes are applied, some parts of the system may attain a negative
internal temperature. This is clearly incorrect, as noted already in
Ref.~\cite{Ripoll_DPDE_1998}, and leads to the simulation instability. The
maximum heat flux $J_{max}$ depends very strongly on $\kappa _{0}$,
especially in the cases where the energy transport by heat conduction dominates 
over the diffusive transport, and
indirectly also on other system parameters, which are more
important in cases when the diffusive transport is high ({\em e.g.}
for low $\kappa _{0}$ values and moderate temperature gradients). The
value of $J_{max}$ appears to be slightly larger for simulations with
$q^A$.
   
\subsection{Energy transfer}
\label{sec:ener}
 
To investigate the mechanisms for heat transfer within a DPDE
fluid in more detail, we define a reference plane to monitor the involved energy
fluxes. This plane is
perpendicular to the gradient direction, and can be placed at different
positions between the hot and cold slabs.  Given a heat flux
$J_0$, energy conservation guarantees
that the total energy flux through a plane at any position is constant, 
$J_{tot}=J_0$. Four different energy
fluxes can be distinguished from two different types.  One type 
is a heat-conduction flux, $J_C$, which corresponds to the exchange of 
internal energy between fluid particles located 
at different sides of the plane within distances smaller than the
cutoff radius $r_{c}$, according to 
the $q^{HC}$ term in Eq.~(\ref{q_E}) or
Eq.~(\ref{q_RE}).
Another contribution is the diffusive flux $J_D$ of energy, 
realized by fluid particles which actually cross the plane.  Mass
conservation enforces that the mass flux through the plane is on
average zero, but due to the externally imposed temperature gradient,
particles on the hot side are more energetic than on the cold side
which results in a diffusive transfer of energy.  In this way,
diffusive energy transfer includes potential, kinetic, and internal
contributions.

\begin{figure}[h]
	\begin{center}
		\includegraphics[width=0.5\textwidth]{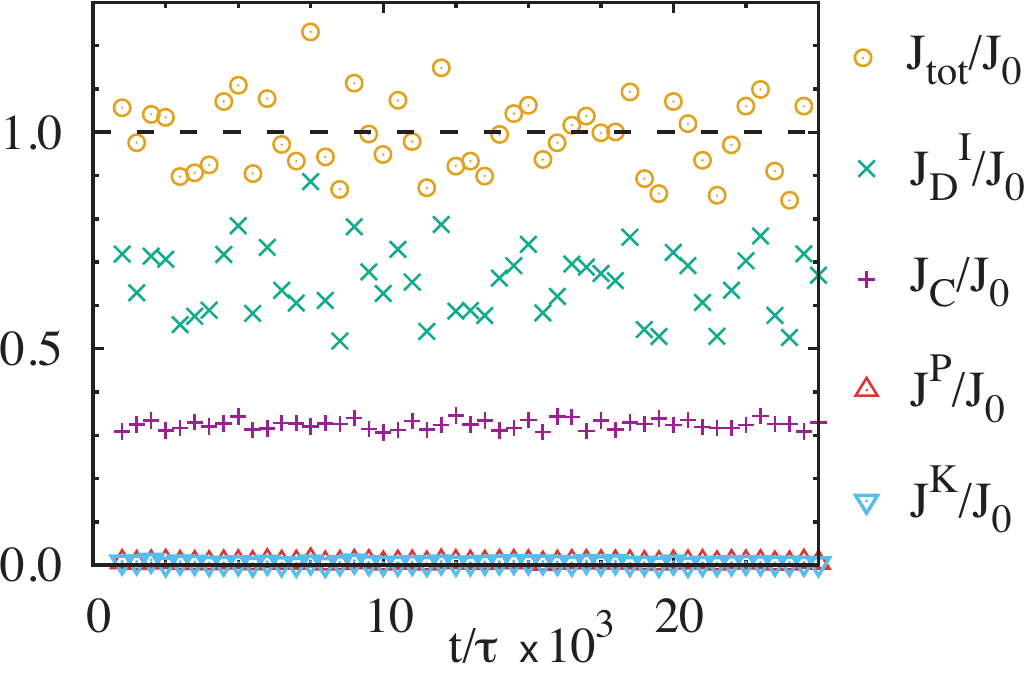}
		\caption{Energy fluxes through
                  a reference plane at $x=8 r_{c}$ (hot area) as a
                  function of time: conductive flux of internal energy
                  $J_C$, and the diffusive fluxes of internal $J_D^I$,
                  potential $J^P$, and kinetic $J^K$ energies. All
                  contributions are normalized by $J_0$, which is
                  $J_0\tau r_c^2/(k_BT_0)=6$ here.}
	\label{fig:local} 	
  \end{center}
\end{figure}

Figure~\ref{fig:local} shows various contributions to the total energy
flux through a plane in the hot area at $x=8 r_{c}$. The major
contributions correspond to the fluxes of particle internal energy,
$J_C$ and $J_D^I$, while the diffusive fluxes from the
potential and kinetic energies, $J^P$ and $J^K$, are very small. This
can be intuitively understood, since the ratio between the particle
internal energy and potential/kinetic energy is proportional to $c_v$,
which is $c_v/k_B = 200$ here. Note that in this example, the
diffusive flux of internal energy $J_D^I$ is larger than the
conductive flux $J_C$, and that $J_{tot}= J_C+J_D$, with
$J_D=J_D^I+J^P+J^K$. Fluctuations of the total flux around the input
value are due to the statistical error of the measurement procedure.

\begin{figure}[h]
	\includegraphics[width=0.47\textwidth]{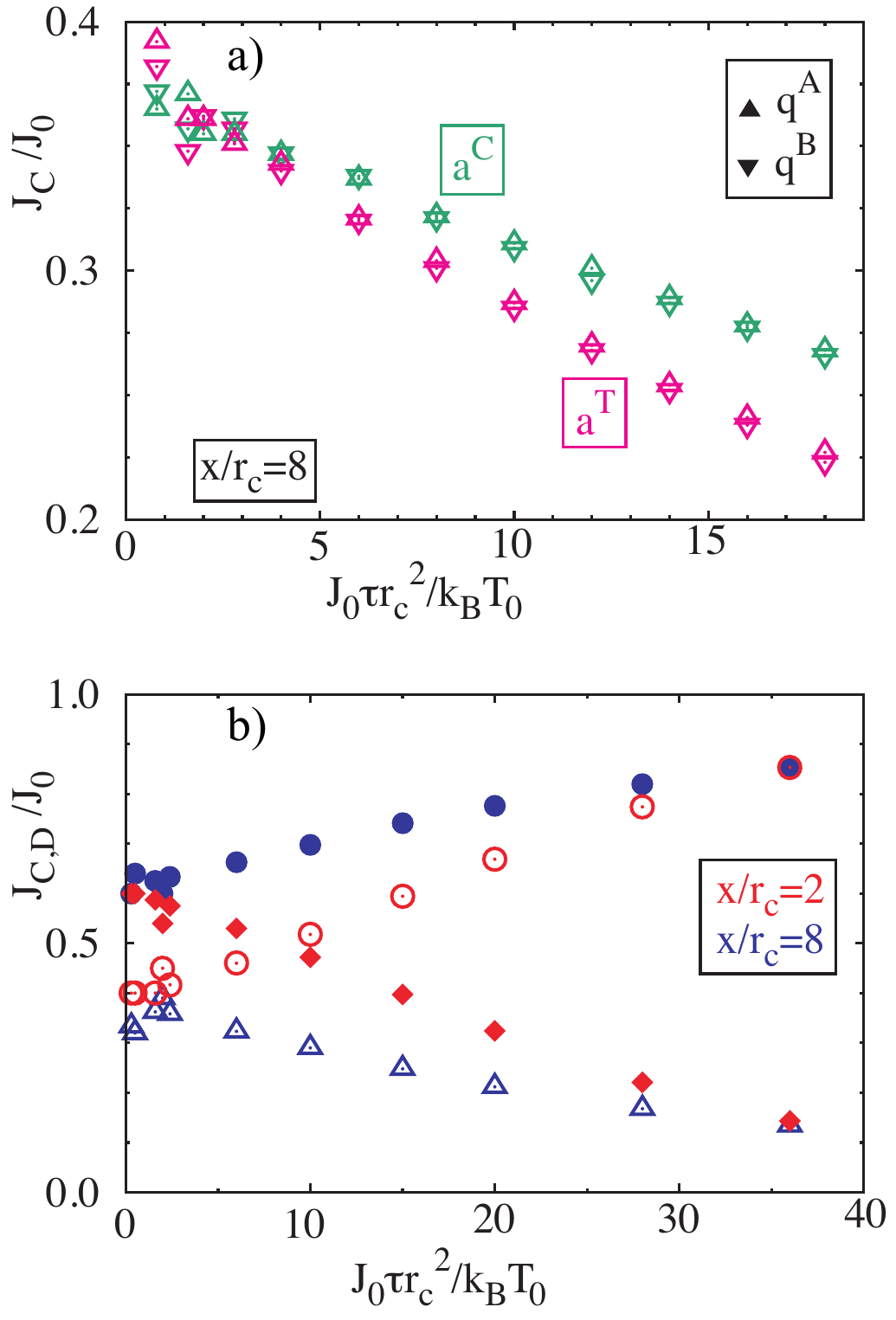}
  \caption{Energy flux across a plane as a function of the
    applied~$J_0$.  
    a)~Conductive energy flux at $x/r_{c} = 8$ for the two heat
    conduction models, $q^A$ in Eq.~(\ref{q_E}) and $q^B$ in
    Eq.~(\ref{q_RE}), and two types of conservative interactions,
    $a^{C}$ and $a^T$. 
    b)~Conductive (open symbols) and diffusive (solid symbols) contributions
    to the energy flux measured through a plane in the cold area
    $x/r_{c} = 2$ (blue), and a plane in the hot area $x/r_{c} = 8$
    (red). At low temperatures, energy flux is
    mainly conductive, while at high temperatures diffusive 
    flux of energy dominates.}
	\label{fig:local_ratio}
\end{figure}

Figure \ref{fig:local_ratio}a shows the conductive flux $J_C$ for
various models through a plane in the hot area at $x/r_{c}=8$
as a function of the externally imposed $J_0$. The differences between
the two implementations of the heat conduction term, $q^A$ and $q^B$,
are systematic, but very small. However, the conductive flux
decays faster for the $a^T$ model of the conservative-force
coefficient in comparison to $a^{C}$ with increasing $J_0$. 
Note that for these two models, the temperature at the reference plane
is the same, but the density is different, 
as can be seen in Fig.~\ref{fig:temp_dens}. 
This monotonic decay of the conductive flux with increasing 
$J_0$ does not occur though, when the reference plane is placed in
the cold area at $x/r_{c} = 2$, as can be seen by the increase of $J_C$ 
(open blue symbols) in Fig.~\ref{fig:local_ratio}b. 
Correspondingly, the diffusive flux $J_D$ increases with the injected
heat through the plane at the hot area, while it decreases through the plane at
the cold area. 
Therefore, local temperature strongly affects not only the overall
heat conductivity as shown in Fig.~\ref{fig:kappa}, but also the ratio
between conductive and diffusive fluxes of energy.  
In Fig.~\ref{fig:local_ratio}b, this means that on the hot side,
the energy flux is mainly diffusive, while on the cold side, it 
is mainly conductive (for not too small applied heat fluxes). 

\begin{figure}[t]
	\includegraphics[width=0.47\textwidth]{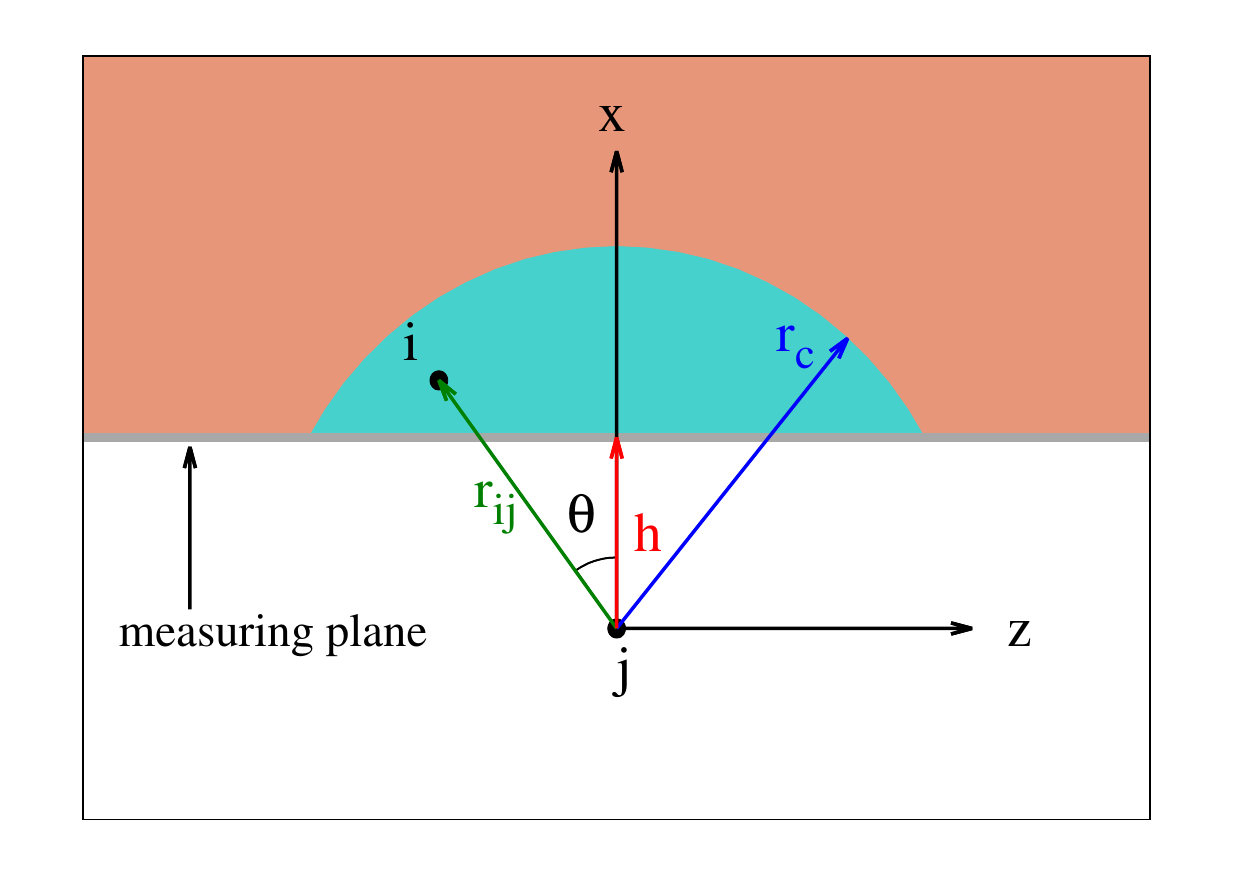}
	\caption{Heat conduction integration domain (cyan) for
		a particle $j$ and particles $i$ at opposite sides of the
		reference plane.\label{integ-heat}}
\end{figure}

\subsection{Analytical calculation of energy transfer}
\label{theory}

In order to provide an analytical
expression for the heat conduction coefficient $\kappa$, we analyze
both the conductive and diffusive fluxes of the internal energy
to total energy transfer, since kinetic and potential energies have a
much smaller contribution (see Fig.~\ref{fig:local}).
The conductive contribution to the energy transfer between two
particles $i$ and $j$ at opposite sides of the reference plane is
given by Eq.~(\ref{q_E}) or Eq.~(\ref{q_RE}). 
These two expressions have been demonstrated to be quantitatively very similar, which means that
the assumption $(T_{i} + T_{j})^2/(T_{i}T_{j}) \simeq 4$ is valid for 
reasonably small temperature differences. The random part of
heat conduction is zero on average, such that the average heat rate from Eq.~(\ref{q_E}) between
two particles at the opposite sides of the reference plane becomes
\begin{equation}
  q_{C,ij}^{A} = \frac{c_{v}^{2}}{k_B}\kappa _{0}^A(T_{j}-T_{i})\omega^{H}(r_{ij}).
\end{equation}
The linear temperature profile can be approximated as
$T_{j}-T_{i} = \dfrac{\mathrm{d}T}{\mathrm{d}x} \Delta x = \dfrac{\mathrm{d}T}{\mathrm{d}x} r_{ij} \cos\theta$, where
$\theta$ is the angle between $r_{ij}$ and the temperature gradient
axis, see Fig.~\ref{integ-heat}.
Considering that particle $j$ is located at a distance $h$ from the
plane, the total heat conduction rate between particle $j$ and its neighbors
on the other side of the reference plane is given by
\begin{equation}\label{eq:qcj}
q_{C,j}^{A}(h) = \rho \int_{0}^{2\pi} \text{d}\phi \int_{0}^{\theta_c}\text{d}\theta 
\int_{h}^{r_{c}}\text{d}r_{ij} r_{ij}^{2}g(r_{ij})\sin\theta  q_{ij}^{A}, 
\end{equation}
with $\theta_c=\arccos (h/r_{c})$. For a non-ideal fluid, the RDF is
$g(r_{ij}) \neq 1$ and needs to be pre-computed for the numerical integration of Eq.~(\ref{eq:qcj}).  
Total heat rate by conduction is then
calculated by integrating $q^A_{C,j}(h)$ over a volume $Ar_c$ as
\begin{equation}
q_{C}^A = \rho A\int_{0}^{r_c} q_{j}^{A}(h)\text{d}h = \frac{\pi \rho^{2}c_{v}^{2}\kappa_{0}^A}{k_B} \dfrac{\mathrm{d}T}{\mathrm{d}x} A H(r_{c}),
\label{eq:h_cond}
\end{equation}
where $A$ is the area of the reference plane and
\begin{eqnarray} \label{eq:h}
H(r_{c}) = & \int_{0}^{r_{c}} \text{d}h \left(1-\dfrac{h^{2}}{r_{c}^{2}}\right) 
\int_{h}^{r_{c}}\text{d}r r^{3}g(r)\omega^{H}(r)  \nonumber \\ 
= & r_c^5 \int_{0}^{1} \text{d}s (1-s^{2}) 
\int_{s}^{1}\text{d}l l^{3}g(lr_c)\omega^{H}(lr_c) \equiv r_c^5 I_1.
\end{eqnarray}
The second equality above makes use of the change of variables $s=h/r_c$ and $l=r/r_c$, while  
the last equality defines a numerical coefficient  $I_1$, which
in general depends on $g(r)$ and $\omega^{H}(r)$, and therefore on the system
parameters.  With the employed default parameters in
Sec.~\ref{sec:pmt}, $I_1 = 0.058$.

\begin{figure}
	\includegraphics[width=0.45\textwidth]{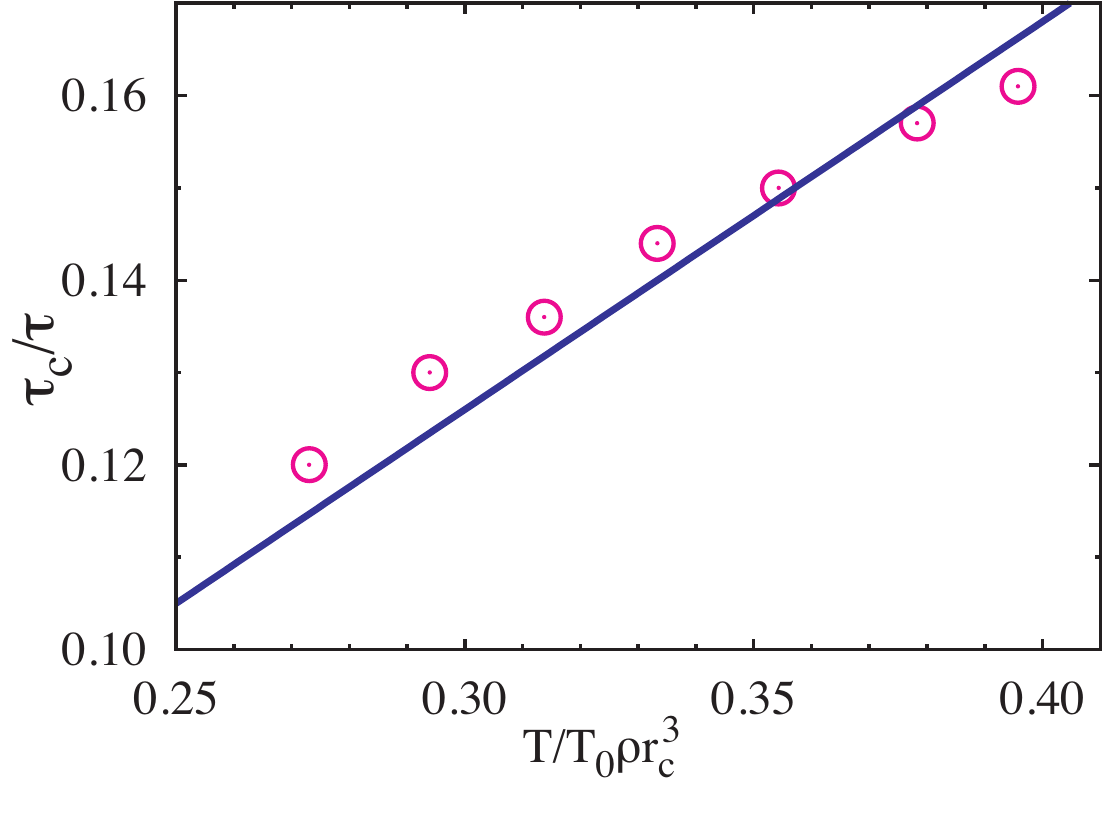}
	\caption{Collision time calculated for various values of the average
		temperature and density. Symbols correspond to simulation results
		and the solid line to Eq.~(\ref{eq:tau}).
		\label{meanfreetime}}
\end{figure}

The second contribution to heat transfer by particle diffusion is
calculated by considering particles crossing the reference plane at $x
=$const. It is equal to $q^I_D = N_J\Delta \epsilon$, where
$N_J$ is the number of particles crossing the reference plane from
hot to cold per unit time.  The internal energy transferred by each particle
crossing from the hot to the cold side can be calculated as
\begin{equation}
\Delta \epsilon = 
\bar{\epsilon}(x+\lambda)-\bar{\epsilon}(x-\lambda) \simeq 
2\lambda \dfrac{\partial\bar{\epsilon}}{\partial T} \dfrac{\mathrm{d} T}{\mathrm{d} x} = 2\lambda c_{v} \dfrac{\mathrm{d} T}{\mathrm{d} x}. 
\end{equation}
Here, the mean free path $\lambda$ can be approximated as $\lambda =
\tau_{c} \bar{v}_{x}$, where $\tau_{c}$ is the collision time and
$\bar{v}_{x}$ is the average velocity of particles crossing the plane in
one direction. The collision time is related to the decay of the
velocity autocorrelation function, which can be calculated in
simulations by assuming an exponential decay of the velocity
autocorrelation as 
\begin{equation}
\left< \mathbf{v}_{i}(t) \mathbf{v}_{i}(0) \right> = \mathrm{e}^{-t/\tau _{c}}\mathbf{v}_{i}^2(0). 
\end{equation}
In order to provide an analytical estimate for 
$\tau_{c}$, we follow the procedure in Ref.~\cite{Groot_DPD_1997}, 
where the friction from dissipative interactions defines $1/ \tau_c$ and 
the sum over dissipative forces for neighboring particles is replaced by 
an integral. The random force vanishes on average and the conservative
force determines the RDF, such that the expression in
Ref.~\cite{Groot_DPD_1997} can be generalized to
\begin{align}\label{eq:tau_inv}
\tau_{c}^{-1}= & \dfrac{4\pi \gamma \rho}{3m} \int_{0}^{r_{c}} \mathrm{d}r r^2 g(r) \omega ^{D}(r) \nonumber \\
= & \dfrac{4\pi \gamma \rho r_c^3}{3m} \int_{0}^{1} \mathrm{d}l l^2 g(l r_c) \omega ^{D}(l r_c) \equiv \dfrac{4\pi \gamma \rho r_c^3}{3m} I_2.
\end{align}
The last equality defines the numerical coefficient $I_2$, which can
be calculated for the default parameters in
Sec.~\ref{sec:pmt} to be $I_2=0.131$.  Substitution of $\gamma =
\sigma^2/(2 k_B T)$ (i.e. for $T_i \simeq T_j$) into
Eq.~(\ref{eq:tau_inv}) results in
\begin{equation}
\label{eq:tau}
\tau_{c} =  \dfrac{3m k_{B}T}{2 \pi \sigma ^{2} \rho r_{c}^{3} I_2}.
\end{equation}
A good agreement between simulated collision times and the analytical approximation 
is shown in Fig.~\ref{meanfreetime} for different values of the average
temperature and density values.

Assuming that the mean free path is small, that the temperature
gradients are not large, and that the velocity distribution function 
$f(v)$ can be approximated by the Maxwell-Boltzmann distribution, we can calculate both 
$N_J$ and $\bar{v}_{x}$. 
The number of particles, $N_J$, crossing the reference plane from hot to cold per unit time, 
accounts only for particles with positive velocities,
\begin{equation}\label{np}
  N_J  =  \rho A \int_{0}^{\infty} f(v_{x}) v_{x}\mathrm{d}v_{x} 
  = \rho A\sqrt{\dfrac{k_{B}T}{2\pi m}},
\end{equation}
and the average velocity of those particles is 
\begin{equation}\label{avevz}
\bar{v}_{x} = \dfrac{\int_{0}^{\infty}\mathrm{d}v_{x} 
v_{x}f(v_{x})}{\int_{0}^{\infty}\mathrm{d}v_{x}f(v_{x})} 
=\sqrt{\dfrac{\pi k_{B}T}{2m}}. 
\end{equation}
The results from Eqs.~(\ref{eq:tau}), (\ref{np}), and (\ref{avevz}) together 
yield the diffusive rate of internal energy
\begin{equation} \label{eq:qdi}
q_D^{I} = \dfrac{3 (k_{B}T)^2 c_v}{2 \pi \sigma ^{2}r_{c}^{3} I_2} A \dfrac{\mathrm{d} T}{\mathrm{d} x}.
\end{equation}
The total heat rate is then the sum of the contributions from Eqs.~(\ref{eq:h_cond}) 
and (\ref{eq:qdi}). With Eq.~(\ref{eq:fourier}), the heat
conductivity can then be approximated as
\begin{equation}\label{con-equ}
\kappa \simeq  \frac{\pi \rho^{2}c_{v}^{2}\kappa_{0}^A r_c^5}{k_B} I_1 + 
\dfrac{3 (k_{B}T)^2 c_v}{2 \pi \sigma ^{2}r_{c}^{3} I_2}. 
\end{equation}
This analytical expression can be compared with the simulation
results shown in Fig.~\ref{fig:kappa}, where a very satisfactory agreement is obtained
without any adjustable parameters. 

\subsection{Prandtl and Schmidt numbers}

A quantitative comparison of the properties of any simulated fluid with
those of real fluids can be achieved by considering dimensionless
numbers, such as the {\em Prandtl number}, $Pr$, or the {\em Schmidt
number}, $Sc$. 
Prandtl number is the ratio of momentum and
energy transport defined as 
\begin{equation}\label{PrNu}
\textrm{Pr}=\dfrac{\mu C_{P}}{\kappa},
\end{equation}
where $\mu$ and $C_{P}$ are dynamic viscosity and specific heat
capacity at constant pressure, respectively. 
For the large values of $c_v$ in DPDE, $C_P \simeq c_v$. 
The Schmidt number $Sc$ is the ratio 
of momentum and diffusive mass transport defined as
\begin{equation}
\textrm{Sc} = \dfrac{\mu}{m\rho D},
\end{equation}
where $D$ is the translational diffusion coefficient.

The Prandtl and Schmidt numbers in DPDE are 
functions of model parameters, and can be set to
various values independently.  The transport properties of a DPD fluid
have been investigated in several
studies~\cite{Groot_DPD_1997,Fan_DNA_2006,Noguchi_TCT_2008}.
Expressions for $D$ and $\mu$ have been derived, and can be generalized in a similar
way as previously shown for the collision time. This yields
\begin{equation} \label{eq:diff}
D = \frac{\tau_c k_B T}{m} = \dfrac{3 (k_{B}T)^2}{2 \pi \sigma ^{2} \rho r_{c}^{3} I_2},
\end{equation}
\begin{align}\label{eq:mu}
\mu = &\dfrac{\rho D}{2} + \dfrac{2\pi \gamma \rho^2}{15} \int_{0}^{r_{c}} \mathrm{d}r r^4 g(r) \omega ^{D}(r)  \nonumber \\
= & \dfrac{\rho D}{2} + \dfrac{\pi \sigma^2 \rho^2 r_c^5}{15 k_B T} \int_{0}^{1} \mathrm{d}l l^4 g(l r_c) \omega ^{D}(l r_c)  \\ 
\equiv & \dfrac{3 (k_{B}T)^2}{4 \pi \sigma ^{2} r_{c}^{3} I_2} + \dfrac{\pi \sigma^2 \rho^2 r_c^5}{15 k_B T} I_3. \nonumber 
\end{align}
Eqution~(\ref{eq:mu}) defines the numerical coefficient $I_3$, which can
be calculated for the default parameters to be $I_3=0.073$, resulting
in $Pr = 2.0$ and $Sc=3.57$.  Alternatively, we obtain slightly larger
values, $Pr = 2.1$ and $Sc=4.1$, by keeping all default parameters but
considering the ideal case with $g(r)=1$.  Furthermore, different
expressions for the weight functions of the conservative, dissipative,
and heat terms will lead to different $Pr$ and $Sc$ values. %
Liquids have typical $Pr$ and $Sc$ values within ranges $5-15$
and $100-1000$, respectively~\cite{Lide_HBC_2004}, while for gases,
both $Pr$ and $Sc$ are typically close to
unity~\cite{Lide_HBC_2004,Kestin_ETP_1984}. Thus, our default system 
lies at the boundary between gaseous and fluid behavior. Using
the expressions for transport coefficients above, model parameters
(e.g. $\rho$, $r_c$, $\sigma$, $c_v$, $\kappa_{0}$) of a DPDE fluid
can be selected such that the Prandtl and Schmidt numbers correspond
more accurately to those of any given fluid. For example, an increase in 
the fluid density $\rho$ and/or the cutoff radius $r_c$ will lead to Prandtl 
and Schmidt numbers which are larger than those employed here, but at 
the same time will result in an increase of computational cost.   

\section{Summary and conclusions}
\label{conlusion}

In this paper, various aspects related to the performance of the DPDE
method have been systematically investigated. A modification of the
velocity-Verlet integration algorithm is suggested, which has a
simple implementation and conserves local and total energies
up to the order of machine precision. In this algorithm, the changes
in local kinetic and potential energies are exactly counterbalanced
by the internal energy every timestep at the level of single DPDE
particles. The model is validated by verifying the equivalence of
the two alternative definitions of the temperature (kinetic and
internal), $T_{K} = T_{I}$, and by studying the behavior of particle
RDF, which suggests that the timestep should be $\Delta t \lesssim
0.02\tau$ (here, $\tau$ is the unit of time) for a stable and 
consistent simulation.

Several choices for the heat-conduction term and conservative
interactions in the DPDE method have been studied and compared. The
interparticle heat conduction in DPDE has been postulated in two
different ways. The expression $q^{B}$ is more intuitive 
since it is directly proportional to the particles' temperature difference 
as expected from the Fourier law, while $q^{A}$ is proportional to the particles' inverse
temperature difference, which has been commonly used so far.
Here, we show that when the prefactors in both approaches are properly
related, the differences between the predictions of both
models are minimal. 
Compressibility effects are especially relevant in the presence of a
temperature gradient, due to the related differences in the density
distribution, such that we investigate different choices for the
conservative interactions. 
A DPDE fluid with temperature-dependent conservative interactions
shows a compressibility lower than a DPDE fluid with an ideal-gas
equation of state, but clearly larger than a DPDE fluid with constant
conservative interactions. 
Nevertheless, the heat conduction coefficients are similar for both 
temperature-dependent and constant conservative interactions.

Thermal conductivity of the DPDE models is measured in simulations by
fitting the Fourier law to an induced temperature gradient. Thermal
conductivity describes the system in a very robust manner, showing
minor changes for a wide range of temperature gradients. 
Energy in a DPDE fluid is transferred by the diffusive motion of DPDE
particles, and by heat conduction due to the internal particle
temperature, which depends on local average temperature. 
The inter-particle conductivity is directly controlled by 
$\kappa_{0}$, while the diffusive transport depends on the
particle friction (or equivalently $\sigma^2$), such that depending
on local conditions, conductive or diffusive transport of heat may
dominate. Good agreement between the analytical expression and
simulation measurements of the thermal conductivity is obtained 
for a large range of parameters. 
We have also presented analytical approaches to obtain the most
important transport coefficients, and therefore the two most relevant
non-dimensional fluid numbers, the Prandtl ($Pr$) and Schmidt ($Sc$)
numbers. These expressions allow the selection of simulation parameters
such that the corresponding $Pr$ and $Sc$ of a DPDE fluid approximate
well those of liquids. In conclusion, our results provide a
detailed guidance on how to properly employ the DPDE method in
simulations of various mesoscopic systems with temperature inhomogeneities.

\appendix

\section{Integration algorithm}
\label{sec:alg}  
  
For integration of Eq.~(\ref{equation_motion}), the velocity-Verlet algorithm~\cite{Allen_CSL_1991} is adapted. In the first integration step, 
both particle velocities and internal energies are advanced half timestep, while particle positions are integrated full timestep as follows
\begin{eqnarray}
\textbf{v}_{i}\left(t+\dfrac{\Delta t}{2}\right) &=& \textbf{v}_{i}(t)+\dfrac{\textbf{F}_{i}(t)}{m_i}\dfrac{\Delta t}{2}, \\
\epsilon_{i}\left(t+\dfrac{\Delta t}{2}\right) &=& \epsilon_{i}(t) + q_{i}(t) \dfrac{\Delta t}{2}   \nonumber \\
&& - \left[ K_{i}\left(t+\dfrac{\Delta t}{2}\right)-K_{i}(t)\right], \\
\textbf{r}_{i}\left(t+\Delta t\right) &=& \textbf{r}_{i}(t)+\textbf{v}_{i}\left(t+\dfrac{\Delta t}{2}\right) \Delta t.
\end{eqnarray}   
Note that the term $[\cdots]$ 
reflects a change in the particle's kinetic energy $K_i(t) = \dfrac{m_i}{2} v_{i}^2(t)$, 
which is counterbalanced by a portion of internal energy in order to conserve the total energy locally. Before the second integration 
step, particle forces $\textbf{F}_i(t+\Delta t)$, potential energies $P_i(t+\Delta t)$, and energy rates $q_{i}(t+\Delta t)$ are computed 
based on $\textbf{r}_i(t+\Delta t)$ and $\textbf{v}_i\left(t+\Delta t/2\right)$. In the second step, particle velocities and internal energies 
are integrated half timestep as  
 \begin{eqnarray}
\textbf{v}_{i}(t+\Delta t) &=& \textbf{v}_{i}\left(t+\dfrac{\Delta t}{2}\right) + \dfrac{\textbf{F}_{i}(t+\Delta t)}{m_i} \dfrac{\Delta t}{2}, \\
\epsilon_{i} (t+\Delta t) &=& \epsilon_{i} \left(t+\dfrac{\Delta t}{2}\right)  + q_{i}(t + \Delta t) \dfrac{\Delta t}{2}  \nonumber \\
&& - \left[K_{i}(t+\Delta t)-K_{i}\left(t+\dfrac{\Delta t}{2}\right)\right] \nonumber \\
&&  - \left[P_{i}(t+\Delta t)-P_{i}(t)\right]. 
\end{eqnarray}
Here, changes in both kinetic and potential energies are offset
against the internal energy to satisfy local energy conservation.
Potential energy corresponds to the conservative interaction as
$\mathbf{F}_{ij}^{C} = -\vec{\nabla} P_{ij}$, such that \mbox{$P_{i}
  = \sum_j\dfrac{a_{ij}r_{c}}{2}\left(1-\dfrac{r_{ij}}{r_{c}}\right)^2$.}

\section*{Acknowledgments}
The authors gratefully acknowledge the computing time granted through JARA-HPC on the supercomputer JURECA \cite{jureca} at Forschungszentrum J\"ulich.

\bibliographystyle{apsrev}
\bibliography{main_new}

\begin{thebibliography}{75}
\expandafter\ifx\csname natexlab\endcsname\relax\def\natexlab#1{#1}\fi
\expandafter\ifx\csname bibnamefont\endcsname\relax
  \def\bibnamefont#1{#1}\fi
\expandafter\ifx\csname bibfnamefont\endcsname\relax
  \def\bibfnamefont#1{#1}\fi
\expandafter\ifx\csname citenamefont\endcsname\relax
  \def\citenamefont#1{#1}\fi
\expandafter\ifx\csname url\endcsname\relax
  \def\url#1{\texttt{#1}}\fi
\expandafter\ifx\csname urlprefix\endcsname\relax\def\urlprefix{URL }\fi
\providecommand{\bibinfo}[2]{#2}
\providecommand{\eprint}[2][]{\url{#2}}

\bibitem[{\citenamefont{Pivkin et~al.}(2011)\citenamefont{Pivkin, Caswell, and
  Karniadakis}}]{Pivkin_DPD_2011}
\bibinfo{author}{\bibfnamefont{I.~V.} \bibnamefont{Pivkin}},
  \bibinfo{author}{\bibfnamefont{B.}~\bibnamefont{Caswell}}, \bibnamefont{and}
  \bibinfo{author}{\bibfnamefont{G.~E.} \bibnamefont{Karniadakis}}, in
  \emph{\bibinfo{booktitle}{Reviews in Computational Chemistry}}, edited by
  \bibinfo{editor}{\bibfnamefont{K.~B.} \bibnamefont{Lipkowitz}}
  (\bibinfo{publisher}{John Wiley \& Sons, Inc.}, \bibinfo{address}{Hoboken,
  NJ, USA}, \bibinfo{year}{2011}), vol.~\bibinfo{volume}{27}, pp.
  \bibinfo{pages}{85--110}.

\bibitem[{\citenamefont{Winkler et~al.}(2014)\citenamefont{Winkler, Fedosov,
  and Gompper}}]{Winkler_DRP_2014}
\bibinfo{author}{\bibfnamefont{R.~G.} \bibnamefont{Winkler}},
  \bibinfo{author}{\bibfnamefont{D.~A.} \bibnamefont{Fedosov}},
  \bibnamefont{and} \bibinfo{author}{\bibfnamefont{G.}~\bibnamefont{Gompper}},
  \bibinfo{journal}{Curr. Opin. Colloid Interface Sci.}
  \textbf{\bibinfo{volume}{19}}, \bibinfo{pages}{594} (\bibinfo{year}{2014}).

\bibitem[{\citenamefont{Fedosov et~al.}(2014)\citenamefont{Fedosov, Noguchi,
  and Gompper}}]{Fedosov_MBF_2014}
\bibinfo{author}{\bibfnamefont{D.~A.} \bibnamefont{Fedosov}},
  \bibinfo{author}{\bibfnamefont{H.}~\bibnamefont{Noguchi}}, \bibnamefont{and}
  \bibinfo{author}{\bibfnamefont{G.}~\bibnamefont{Gompper}},
  \bibinfo{journal}{Biomech. Model. Mechanobiol.}
  \textbf{\bibinfo{volume}{13}}, \bibinfo{pages}{239} (\bibinfo{year}{2014}).

\bibitem[{\citenamefont{Elgeti et~al.}(2015)\citenamefont{Elgeti, Winkler, and
  Gompper}}]{Elgeti_PMS_2015}
\bibinfo{author}{\bibfnamefont{J.}~\bibnamefont{Elgeti}},
  \bibinfo{author}{\bibfnamefont{R.~G.} \bibnamefont{Winkler}},
  \bibnamefont{and} \bibinfo{author}{\bibfnamefont{G.}~\bibnamefont{Gompper}},
  \bibinfo{journal}{Rep. Prog. Phys.} \textbf{\bibinfo{volume}{78}},
  \bibinfo{pages}{056601} (\bibinfo{year}{2015}).

\bibitem[{\citenamefont{Schiller et~al.}(2018)\citenamefont{Schiller,
  Kr{\"u}ger, and Henrich}}]{Schiller_MMS_2018}
\bibinfo{author}{\bibfnamefont{U.~D.} \bibnamefont{Schiller}},
  \bibinfo{author}{\bibfnamefont{T.}~\bibnamefont{Kr{\"u}ger}},
  \bibnamefont{and} \bibinfo{author}{\bibfnamefont{O.}~\bibnamefont{Henrich}},
  \bibinfo{journal}{Soft Matter} \textbf{\bibinfo{volume}{14}},
  \bibinfo{pages}{9} (\bibinfo{year}{2018}).

\bibitem[{\citenamefont{Gompper et~al.}(2009)\citenamefont{Gompper, Ihle,
  Kroll, and Winkler}}]{Gompper_APS_2009}
\bibinfo{author}{\bibfnamefont{G.}~\bibnamefont{Gompper}},
  \bibinfo{author}{\bibfnamefont{T.}~\bibnamefont{Ihle}},
  \bibinfo{author}{\bibfnamefont{D.~M.} \bibnamefont{Kroll}}, \bibnamefont{and}
  \bibinfo{author}{\bibfnamefont{R.~G.} \bibnamefont{Winkler}},
  \bibinfo{journal}{Adv. Polym. Sci.} \textbf{\bibinfo{volume}{221}},
  \bibinfo{pages}{1} (\bibinfo{year}{2009}).

\bibitem[{\citenamefont{McNamara and Zanetti}(1988)}]{McNamara_UBE_1988}
\bibinfo{author}{\bibfnamefont{G.~R.} \bibnamefont{McNamara}} \bibnamefont{and}
  \bibinfo{author}{\bibfnamefont{G.}~\bibnamefont{Zanetti}},
  \bibinfo{journal}{Phys. Rev. Lett.} \textbf{\bibinfo{volume}{61}},
  \bibinfo{pages}{2332} (\bibinfo{year}{1988}).

\bibitem[{\citenamefont{He and Luo}(1997)}]{He_LBM_1997}
\bibinfo{author}{\bibfnamefont{X.}~\bibnamefont{He}} \bibnamefont{and}
  \bibinfo{author}{\bibfnamefont{L.-S.} \bibnamefont{Luo}},
  \bibinfo{journal}{Phys. Rev. E} \textbf{\bibinfo{volume}{56}},
  \bibinfo{pages}{6811} (\bibinfo{year}{1997}).

\bibitem[{\citenamefont{Succi}(2001)}]{Succi_LBM_2001}
\bibinfo{author}{\bibfnamefont{S.}~\bibnamefont{Succi}},
  \emph{\bibinfo{title}{The {L}attice {B}oltzmann equation for fluid dynamics
  and beyond}} (\bibinfo{publisher}{Oxford University Press},
  \bibinfo{address}{Oxford}, \bibinfo{year}{2001}).

\bibitem[{\citenamefont{Malevanets and Kapral}(1999)}]{Malevanets_MSM_1999}
\bibinfo{author}{\bibfnamefont{A.}~\bibnamefont{Malevanets}} \bibnamefont{and}
  \bibinfo{author}{\bibfnamefont{R.}~\bibnamefont{Kapral}},
  \bibinfo{journal}{J. Chem. Phys.} \textbf{\bibinfo{volume}{110}},
  \bibinfo{pages}{8605} (\bibinfo{year}{1999}).

\bibitem[{\citenamefont{Malevanets and Kapral}(2000)}]{Malevanets_SMD_2000}
\bibinfo{author}{\bibfnamefont{A.}~\bibnamefont{Malevanets}} \bibnamefont{and}
  \bibinfo{author}{\bibfnamefont{R.}~\bibnamefont{Kapral}},
  \bibinfo{journal}{J. Chem. Phys.} \textbf{\bibinfo{volume}{112}},
  \bibinfo{pages}{7260} (\bibinfo{year}{2000}).

\bibitem[{\citenamefont{Kapral}(2008)}]{Kapral_ACP_2008}
\bibinfo{author}{\bibfnamefont{R.}~\bibnamefont{Kapral}},
  \bibinfo{journal}{Adv. Chem. Phys.} \textbf{\bibinfo{volume}{140}},
  \bibinfo{pages}{89} (\bibinfo{year}{2008}).

\bibitem[{\citenamefont{Hoogerbrugge and
  Koelman}(1992)}]{Hoogerbrugge_SMH_1992}
\bibinfo{author}{\bibfnamefont{P.~J.} \bibnamefont{Hoogerbrugge}}
  \bibnamefont{and} \bibinfo{author}{\bibfnamefont{J.~M. V.~A.}
  \bibnamefont{Koelman}}, \bibinfo{journal}{Europhys. Lett.}
  \textbf{\bibinfo{volume}{19}}, \bibinfo{pages}{155} (\bibinfo{year}{1992}).

\bibitem[{\citenamefont{Espa\~{n}ol and Warren}(1995)}]{Espanol_SMO_1995}
\bibinfo{author}{\bibfnamefont{P.}~\bibnamefont{Espa\~{n}ol}} \bibnamefont{and}
  \bibinfo{author}{\bibfnamefont{P.}~\bibnamefont{Warren}},
  \bibinfo{journal}{Europhys. Lett.} \textbf{\bibinfo{volume}{30}},
  \bibinfo{pages}{191} (\bibinfo{year}{1995}).

\bibitem[{\citenamefont{Espa\~{n}ol and Warren}(2017)}]{Espanol_DPD_2017}
\bibinfo{author}{\bibfnamefont{P.}~\bibnamefont{Espa\~{n}ol}} \bibnamefont{and}
  \bibinfo{author}{\bibfnamefont{P.~B.} \bibnamefont{Warren}},
  \bibinfo{journal}{J. Chem. Phys.} \textbf{\bibinfo{volume}{146}},
  \bibinfo{pages}{150901} (\bibinfo{year}{2017}).

\bibitem[{\citenamefont{Cleary}(1998)}]{Cleary_MCF_1998}
\bibinfo{author}{\bibfnamefont{P.~W.} \bibnamefont{Cleary}},
  \bibinfo{journal}{Appl. Math. Model.} \textbf{\bibinfo{volume}{22}},
  \bibinfo{pages}{981} (\bibinfo{year}{1998}).

\bibitem[{\citenamefont{Sarman and Evans}(1992)}]{Sarman_HFM_1992}
\bibinfo{author}{\bibfnamefont{S.}~\bibnamefont{Sarman}} \bibnamefont{and}
  \bibinfo{author}{\bibfnamefont{D.~J.} \bibnamefont{Evans}},
  \bibinfo{journal}{Phys. Rev. A} \textbf{\bibinfo{volume}{45}},
  \bibinfo{pages}{2370} (\bibinfo{year}{1992}).

\bibitem[{\citenamefont{Wold and Hafskjold}(1999)}]{Wold_NMD_1999}
\bibinfo{author}{\bibfnamefont{I.}~\bibnamefont{Wold}} \bibnamefont{and}
  \bibinfo{author}{\bibfnamefont{B.}~\bibnamefont{Hafskjold}},
  \bibinfo{journal}{Int. J. Thermophys.} \textbf{\bibinfo{volume}{20}},
  \bibinfo{pages}{847} (\bibinfo{year}{1999}).

\bibitem[{\citenamefont{Reith and M{\"u}ller-Plathe}(2000)}]{Reith_NTD_2000}
\bibinfo{author}{\bibfnamefont{D.}~\bibnamefont{Reith}} \bibnamefont{and}
  \bibinfo{author}{\bibfnamefont{F.}~\bibnamefont{M{\"u}ller-Plathe}},
  \bibinfo{journal}{J. Chem. Phys.} \textbf{\bibinfo{volume}{112}},
  \bibinfo{pages}{2436} (\bibinfo{year}{2000}).

\bibitem[{\citenamefont{Artola et~al.}(2008)\citenamefont{Artola, Rousseau, and
  Galliero}}]{Artola_MTD_2008}
\bibinfo{author}{\bibfnamefont{P.-A.} \bibnamefont{Artola}},
  \bibinfo{author}{\bibfnamefont{B.}~\bibnamefont{Rousseau}}, \bibnamefont{and}
  \bibinfo{author}{\bibfnamefont{G.}~\bibnamefont{Galliero}},
  \bibinfo{journal}{J. Am. Chem. Soc.} \textbf{\bibinfo{volume}{130}},
  \bibinfo{pages}{10963} (\bibinfo{year}{2008}).

\bibitem[{\citenamefont{Galliero and Volz}(2008)}]{Galliero_TMD_2008}
\bibinfo{author}{\bibfnamefont{G.}~\bibnamefont{Galliero}} \bibnamefont{and}
  \bibinfo{author}{\bibfnamefont{S.}~\bibnamefont{Volz}}, \bibinfo{journal}{J.
  Chem. Phys.} \textbf{\bibinfo{volume}{128}}, \bibinfo{pages}{064505}
  (\bibinfo{year}{2008}).

\bibitem[{\citenamefont{Volpe et~al.}(2011)\citenamefont{Volpe, Buttinoni,
  Vogt, K\"ummerer, and Bechinger}}]{Volpe_MSE_2011}
\bibinfo{author}{\bibfnamefont{G.}~\bibnamefont{Volpe}},
  \bibinfo{author}{\bibfnamefont{I.}~\bibnamefont{Buttinoni}},
  \bibinfo{author}{\bibfnamefont{D.}~\bibnamefont{Vogt}},
  \bibinfo{author}{\bibfnamefont{H.-J.} \bibnamefont{K\"ummerer}},
  \bibnamefont{and}
  \bibinfo{author}{\bibfnamefont{C.}~\bibnamefont{Bechinger}},
  \bibinfo{journal}{Soft Matter} \textbf{\bibinfo{volume}{7}},
  \bibinfo{pages}{8810} (\bibinfo{year}{2011}).

\bibitem[{\citenamefont{Fedosov et~al.}(2015)\citenamefont{Fedosov, Sengupta,
  and Gompper}}]{Fedosov_EFC_2015}
\bibinfo{author}{\bibfnamefont{D.~A.} \bibnamefont{Fedosov}},
  \bibinfo{author}{\bibfnamefont{A.}~\bibnamefont{Sengupta}}, \bibnamefont{and}
  \bibinfo{author}{\bibfnamefont{G.}~\bibnamefont{Gompper}},
  \bibinfo{journal}{Soft Matter} \textbf{\bibinfo{volume}{11}},
  \bibinfo{pages}{6703} (\bibinfo{year}{2015}).

\bibitem[{\citenamefont{L\"usebrink et~al.}(2012)\citenamefont{L\"usebrink,
  Yang, and Ripoll}}]{Luesebrink_TPC_2012}
\bibinfo{author}{\bibfnamefont{D.}~\bibnamefont{L\"usebrink}},
  \bibinfo{author}{\bibfnamefont{M.}~\bibnamefont{Yang}}, \bibnamefont{and}
  \bibinfo{author}{\bibfnamefont{M.}~\bibnamefont{Ripoll}},
  \bibinfo{journal}{J. Phys.: Condens. Matter} \textbf{\bibinfo{volume}{24}},
  \bibinfo{pages}{284132} (\bibinfo{year}{2012}).

\bibitem[{\citenamefont{Tan et~al.}(2017)\citenamefont{Tan, Yang, and
  Ripoll}}]{Tan_ATT_2017}
\bibinfo{author}{\bibfnamefont{Z.}~\bibnamefont{Tan}},
  \bibinfo{author}{\bibfnamefont{M.}~\bibnamefont{Yang}}, \bibnamefont{and}
  \bibinfo{author}{\bibfnamefont{M.}~\bibnamefont{Ripoll}},
  \bibinfo{journal}{Soft Matter} \textbf{\bibinfo{volume}{13}},
  \bibinfo{pages}{7283} (\bibinfo{year}{2017}).

\bibitem[{\citenamefont{Sellan et~al.}(2010)\citenamefont{Sellan, Landry,
  Turney, McGaughey, and Amon}}]{Sellan_EMD_2010}
\bibinfo{author}{\bibfnamefont{D.~P.} \bibnamefont{Sellan}},
  \bibinfo{author}{\bibfnamefont{E.~S.} \bibnamefont{Landry}},
  \bibinfo{author}{\bibfnamefont{J.~E.} \bibnamefont{Turney}},
  \bibinfo{author}{\bibfnamefont{A.~J.~H.} \bibnamefont{McGaughey}},
  \bibnamefont{and} \bibinfo{author}{\bibfnamefont{C.~H.} \bibnamefont{Amon}},
  \bibinfo{journal}{Phys. Rev. B} \textbf{\bibinfo{volume}{81}},
  \bibinfo{pages}{214305} (\bibinfo{year}{2010}).

\bibitem[{\citenamefont{McNamara et~al.}(1997)\citenamefont{McNamara, Garcia,
  and Alder}}]{McNamara_HDC_1997}
\bibinfo{author}{\bibfnamefont{G.~R.} \bibnamefont{McNamara}},
  \bibinfo{author}{\bibfnamefont{A.~L.} \bibnamefont{Garcia}},
  \bibnamefont{and} \bibinfo{author}{\bibfnamefont{B.~J.} \bibnamefont{Alder}},
  \bibinfo{journal}{J. Stat. Phys.} \textbf{\bibinfo{volume}{87}},
  \bibinfo{pages}{1111} (\bibinfo{year}{1997}).

\bibitem[{\citenamefont{Shan}(1997)}]{Shan_SRB_1997}
\bibinfo{author}{\bibfnamefont{X.}~\bibnamefont{Shan}}, \bibinfo{journal}{Phys.
  Rev. E} \textbf{\bibinfo{volume}{55}}, \bibinfo{pages}{2780}
  (\bibinfo{year}{1997}).

\bibitem[{\citenamefont{Yang and Ripoll}(2011)}]{Yang_TNS_2011}
\bibinfo{author}{\bibfnamefont{M.}~\bibnamefont{Yang}} \bibnamefont{and}
  \bibinfo{author}{\bibfnamefont{M.}~\bibnamefont{Ripoll}},
  \bibinfo{journal}{Phys. Rev. E} \textbf{\bibinfo{volume}{84}},
  \bibinfo{pages}{061401} (\bibinfo{year}{2011}).

\bibitem[{\citenamefont{Yang et~al.}(2014)\citenamefont{Yang, Liu, Ripoll, and
  Chen}}]{Yang_MTP_2014}
\bibinfo{author}{\bibfnamefont{M.}~\bibnamefont{Yang}},
  \bibinfo{author}{\bibfnamefont{R.}~\bibnamefont{Liu}},
  \bibinfo{author}{\bibfnamefont{M.}~\bibnamefont{Ripoll}}, \bibnamefont{and}
  \bibinfo{author}{\bibfnamefont{K.}~\bibnamefont{Chen}},
  \bibinfo{journal}{Nanoscale} \textbf{\bibinfo{volume}{6}},
  \bibinfo{pages}{13550} (\bibinfo{year}{2014}).

\bibitem[{\citenamefont{Yang and Ripoll}(2016)}]{Yang_TOM_2016}
\bibinfo{author}{\bibfnamefont{M.}~\bibnamefont{Yang}} \bibnamefont{and}
  \bibinfo{author}{\bibfnamefont{M.}~\bibnamefont{Ripoll}},
  \bibinfo{journal}{Soft Matter} \textbf{\bibinfo{volume}{12}},
  \bibinfo{pages}{8564} (\bibinfo{year}{2016}).

\bibitem[{\citenamefont{Langenberg and M{\"u}ller}(2016)}]{Langenberg_eMC_2016}
\bibinfo{author}{\bibfnamefont{M.}~\bibnamefont{Langenberg}} \bibnamefont{and}
  \bibinfo{author}{\bibfnamefont{M.}~\bibnamefont{M{\"u}ller}},
  \bibinfo{journal}{Europhys. Lett.} \textbf{\bibinfo{volume}{114}},
  \bibinfo{pages}{20001} (\bibinfo{year}{2016}).

\bibitem[{\citenamefont{Espa\~{n}ol}(1997)}]{Espanol_DPDE_1997}
\bibinfo{author}{\bibfnamefont{P.}~\bibnamefont{Espa\~{n}ol}},
  \bibinfo{journal}{Europhys. Lett.} \textbf{\bibinfo{volume}{40}},
  \bibinfo{pages}{631} (\bibinfo{year}{1997}).

\bibitem[{\citenamefont{{Bonet Avalos} and Mackie}(1997)}]{Avalos_DPDE_1997}
\bibinfo{author}{\bibfnamefont{J.}~\bibnamefont{{Bonet Avalos}}}
  \bibnamefont{and} \bibinfo{author}{\bibfnamefont{A.~D.}
  \bibnamefont{Mackie}}, \bibinfo{journal}{Europhys. Lett.}
  \textbf{\bibinfo{volume}{40}}, \bibinfo{pages}{141} (\bibinfo{year}{1997}).

\bibitem[{\citenamefont{Abu-Nada}(2010)}]{AbuNada_CHT_2010}
\bibinfo{author}{\bibfnamefont{E.}~\bibnamefont{Abu-Nada}},
  \bibinfo{journal}{Phys. Rev. E} \textbf{\bibinfo{volume}{81}},
  \bibinfo{pages}{056704} (\bibinfo{year}{2010}).

\bibitem[{\citenamefont{Yamada et~al.}(2011)\citenamefont{Yamada, Kumar, Asako,
  Gregory, and Faghri}}]{Yamada_CHT_2011}
\bibinfo{author}{\bibfnamefont{T.}~\bibnamefont{Yamada}},
  \bibinfo{author}{\bibfnamefont{A.}~\bibnamefont{Kumar}},
  \bibinfo{author}{\bibfnamefont{Y.}~\bibnamefont{Asako}},
  \bibinfo{author}{\bibfnamefont{O.~J.} \bibnamefont{Gregory}},
  \bibnamefont{and} \bibinfo{author}{\bibfnamefont{M.}~\bibnamefont{Faghri}},
  \bibinfo{journal}{Numer. Heat Tr. A-Appl.} \textbf{\bibinfo{volume}{60}},
  \bibinfo{pages}{651} (\bibinfo{year}{2011}).

\bibitem[{\citenamefont{Li et~al.}(2014)\citenamefont{Li, Tang, Lei, Caswell,
  and Karniadakis}}]{Li_EDPD_2014}
\bibinfo{author}{\bibfnamefont{Z.}~\bibnamefont{Li}},
  \bibinfo{author}{\bibfnamefont{Y.-H.} \bibnamefont{Tang}},
  \bibinfo{author}{\bibfnamefont{H.}~\bibnamefont{Lei}},
  \bibinfo{author}{\bibfnamefont{B.}~\bibnamefont{Caswell}}, \bibnamefont{and}
  \bibinfo{author}{\bibfnamefont{G.~E.} \bibnamefont{Karniadakis}},
  \bibinfo{journal}{J. Comp. Phys.} \textbf{\bibinfo{volume}{265}},
  \bibinfo{pages}{113} (\bibinfo{year}{2014}).

\bibitem[{\citenamefont{Yamada et~al.}(2016)\citenamefont{Yamada, Johansson,
  Sund{\'e}n, and Yuan}}]{Yamada_DPD_2016}
\bibinfo{author}{\bibfnamefont{T.}~\bibnamefont{Yamada}},
  \bibinfo{author}{\bibfnamefont{E.~O.} \bibnamefont{Johansson}},
  \bibinfo{author}{\bibfnamefont{B.}~\bibnamefont{Sund{\'e}n}},
  \bibnamefont{and} \bibinfo{author}{\bibfnamefont{J.}~\bibnamefont{Yuan}},
  \bibinfo{journal}{Numer. Heat Tr. A-Appl.} \textbf{\bibinfo{volume}{70}},
  \bibinfo{pages}{595} (\bibinfo{year}{2016}).

\bibitem[{\citenamefont{{Bonet Avalos} and Mackie}(1999)}]{Avalos_DPDE_1999}
\bibinfo{author}{\bibfnamefont{J.}~\bibnamefont{{Bonet Avalos}}}
  \bibnamefont{and} \bibinfo{author}{\bibfnamefont{A.~D.}
  \bibnamefont{Mackie}}, \bibinfo{journal}{J. Chem. Phys.}
  \textbf{\bibinfo{volume}{111}}, \bibinfo{pages}{5267} (\bibinfo{year}{1999}).

\bibitem[{\citenamefont{Mackie et~al.}(1999)\citenamefont{Mackie, {Bonet
  Avalos}, and Navas}}]{Mackie_DPDE_1999}
\bibinfo{author}{\bibfnamefont{A.~D.} \bibnamefont{Mackie}},
  \bibinfo{author}{\bibfnamefont{J.}~\bibnamefont{{Bonet Avalos}}},
  \bibnamefont{and} \bibinfo{author}{\bibfnamefont{V.}~\bibnamefont{Navas}},
  \bibinfo{journal}{Phys. Chem. Chem. Phys.} \textbf{\bibinfo{volume}{1}},
  \bibinfo{pages}{2039} (\bibinfo{year}{1999}).

\bibitem[{\citenamefont{L\'isal et~al.}(2011)\citenamefont{L\'isal, Brennan,
  and {Bonet Avalos}}}]{Lisal_DPD_2011}
\bibinfo{author}{\bibfnamefont{M.}~\bibnamefont{L\'isal}},
  \bibinfo{author}{\bibfnamefont{J.~K.} \bibnamefont{Brennan}},
  \bibnamefont{and} \bibinfo{author}{\bibfnamefont{J.}~\bibnamefont{{Bonet
  Avalos}}}, \bibinfo{journal}{J. Chem. Phys.} \textbf{\bibinfo{volume}{135}},
  \bibinfo{pages}{204105} (\bibinfo{year}{2011}).

\bibitem[{\citenamefont{Larentzos et~al.}(2014)\citenamefont{Larentzos,
  Brennan, Moore, L{\'\i}sal, and Mattson}}]{Larentzos_DPDE_2014}
\bibinfo{author}{\bibfnamefont{J.~P.} \bibnamefont{Larentzos}},
  \bibinfo{author}{\bibfnamefont{J.~K.} \bibnamefont{Brennan}},
  \bibinfo{author}{\bibfnamefont{J.~D.} \bibnamefont{Moore}},
  \bibinfo{author}{\bibfnamefont{M.}~\bibnamefont{L{\'\i}sal}},
  \bibnamefont{and} \bibinfo{author}{\bibfnamefont{W.~D.}
  \bibnamefont{Mattson}}, \bibinfo{journal}{Comput. Phys. Commun.}
  \textbf{\bibinfo{volume}{185}}, \bibinfo{pages}{1987} (\bibinfo{year}{2014}).

\bibitem[{\citenamefont{Homman et~al.}(2016)\citenamefont{Homman, Maillet,
  Roussel, and Stoltz}}]{Homman_DPDE_2016}
\bibinfo{author}{\bibfnamefont{A.-A.} \bibnamefont{Homman}},
  \bibinfo{author}{\bibfnamefont{J.-B.} \bibnamefont{Maillet}},
  \bibinfo{author}{\bibfnamefont{J.}~\bibnamefont{Roussel}}, \bibnamefont{and}
  \bibinfo{author}{\bibfnamefont{G.}~\bibnamefont{Stoltz}},
  \bibinfo{journal}{J. Chem. Phys.} \textbf{\bibinfo{volume}{144}},
  \bibinfo{pages}{024112} (\bibinfo{year}{2016}).

\bibitem[{\citenamefont{Marsh et~al.}(1997)\citenamefont{Marsh, Backx, and
  Ernst}}]{Marsh_FPB_1997}
\bibinfo{author}{\bibfnamefont{C.~A.} \bibnamefont{Marsh}},
  \bibinfo{author}{\bibfnamefont{G.}~\bibnamefont{Backx}}, \bibnamefont{and}
  \bibinfo{author}{\bibfnamefont{M.~H.} \bibnamefont{Ernst}},
  \bibinfo{journal}{Europhys. Lett.} \textbf{\bibinfo{volume}{38}},
  \bibinfo{pages}{411} (\bibinfo{year}{1997}).

\bibitem[{\citenamefont{Ripoll et~al.}(1998)\citenamefont{Ripoll, Espa{\~n}ol,
  and Ernst}}]{Ripoll_DPDE_1998}
\bibinfo{author}{\bibfnamefont{M.}~\bibnamefont{Ripoll}},
  \bibinfo{author}{\bibfnamefont{P.}~\bibnamefont{Espa{\~n}ol}},
  \bibnamefont{and} \bibinfo{author}{\bibfnamefont{M.~H.} \bibnamefont{Ernst}},
  \bibinfo{journal}{Int. J. Mod. Phys. C} \textbf{\bibinfo{volume}{9}},
  \bibinfo{pages}{1329} (\bibinfo{year}{1998}).

\bibitem[{\citenamefont{Qiao and He}(2007)}]{Qiao_SHC_2007}
\bibinfo{author}{\bibfnamefont{R.}~\bibnamefont{Qiao}} \bibnamefont{and}
  \bibinfo{author}{\bibfnamefont{P.}~\bibnamefont{He}}, \bibinfo{journal}{Mol.
  Sim.} \textbf{\bibinfo{volume}{33}}, \bibinfo{pages}{677}
  (\bibinfo{year}{2007}).

\bibitem[{\citenamefont{Marsh and Coveney}(1998)}]{Marsh_DBH_1998}
\bibinfo{author}{\bibfnamefont{C.~A.} \bibnamefont{Marsh}} \bibnamefont{and}
  \bibinfo{author}{\bibfnamefont{P.~V.} \bibnamefont{Coveney}},
  \bibinfo{journal}{J. Phys. A} \textbf{\bibinfo{volume}{31}},
  \bibinfo{pages}{6561} (\bibinfo{year}{1998}).

\bibitem[{\citenamefont{Ripoll and
  Ernst}(2005{\natexlab{a}})}]{Ripoll_MSF_2005}
\bibinfo{author}{\bibfnamefont{M.}~\bibnamefont{Ripoll}} \bibnamefont{and}
  \bibinfo{author}{\bibfnamefont{M.~H.} \bibnamefont{Ernst}},
  \bibinfo{journal}{Phys. Rev. E} \textbf{\bibinfo{volume}{71}},
  \bibinfo{pages}{041104} (\bibinfo{year}{2005}{\natexlab{a}}).

\bibitem[{\citenamefont{Ripoll and
  Ernst}(2005{\natexlab{b}})}]{Ripoll_PLT_2005}
\bibinfo{author}{\bibfnamefont{M.}~\bibnamefont{Ripoll}} \bibnamefont{and}
  \bibinfo{author}{\bibfnamefont{M.~H.} \bibnamefont{Ernst}},
  \bibinfo{journal}{Phys. Rev. E} \textbf{\bibinfo{volume}{72}},
  \bibinfo{pages}{011101} (\bibinfo{year}{2005}{\natexlab{b}}).

\bibitem[{\citenamefont{Shardlow}(2003)}]{Shardlow_DPD_2003}
\bibinfo{author}{\bibfnamefont{T.}~\bibnamefont{Shardlow}},
  \bibinfo{journal}{SIAM J. Sci. Comput.} \textbf{\bibinfo{volume}{24}},
  \bibinfo{pages}{1267} (\bibinfo{year}{2003}).

\bibitem[{\citenamefont{{De Fabritiis} et~al.}(2006)\citenamefont{{De
  Fabritiis}, Serrano, Espa{\~n}ol, and Coveney}}]{De_Fabritiis_ENI_2006}
\bibinfo{author}{\bibfnamefont{G.}~\bibnamefont{{De Fabritiis}}},
  \bibinfo{author}{\bibfnamefont{M.}~\bibnamefont{Serrano}},
  \bibinfo{author}{\bibfnamefont{P.}~\bibnamefont{Espa{\~n}ol}},
  \bibnamefont{and} \bibinfo{author}{\bibfnamefont{P.~V.}
  \bibnamefont{Coveney}}, \bibinfo{journal}{Physica A}
  \textbf{\bibinfo{volume}{361}}, \bibinfo{pages}{429} (\bibinfo{year}{2006}).

\bibitem[{\citenamefont{Serrano et~al.}(2006)\citenamefont{Serrano, {De
  Fabritiis}, Espa{\~n}ol, and Coveney}}]{Serrano_STA_2006}
\bibinfo{author}{\bibfnamefont{M.}~\bibnamefont{Serrano}},
  \bibinfo{author}{\bibfnamefont{G.}~\bibnamefont{{De Fabritiis}}},
  \bibinfo{author}{\bibfnamefont{P.}~\bibnamefont{Espa{\~n}ol}},
  \bibnamefont{and} \bibinfo{author}{\bibfnamefont{P.~V.}
  \bibnamefont{Coveney}}, \bibinfo{journal}{Math. Comput. Simul.}
  \textbf{\bibinfo{volume}{72}}, \bibinfo{pages}{190} (\bibinfo{year}{2006}).

\bibitem[{\citenamefont{Fan et~al.}(2006)\citenamefont{Fan, Phan-Thien, Chen,
  Wu, and Ng}}]{Fan_DNA_2006}
\bibinfo{author}{\bibfnamefont{X.}~\bibnamefont{Fan}},
  \bibinfo{author}{\bibfnamefont{N.}~\bibnamefont{Phan-Thien}},
  \bibinfo{author}{\bibfnamefont{S.}~\bibnamefont{Chen}},
  \bibinfo{author}{\bibfnamefont{X.}~\bibnamefont{Wu}}, \bibnamefont{and}
  \bibinfo{author}{\bibfnamefont{T.~Y.} \bibnamefont{Ng}},
  \bibinfo{journal}{Phys. Fluids} \textbf{\bibinfo{volume}{18}},
  \bibinfo{pages}{063102} (\bibinfo{year}{2006}).

\bibitem[{\citenamefont{Fedosov et~al.}(2008)\citenamefont{Fedosov, Pivkin, and
  Karniadakis}}]{Fedosov_VL_2008}
\bibinfo{author}{\bibfnamefont{D.~A.} \bibnamefont{Fedosov}},
  \bibinfo{author}{\bibfnamefont{I.~V.} \bibnamefont{Pivkin}},
  \bibnamefont{and} \bibinfo{author}{\bibfnamefont{G.~E.}
  \bibnamefont{Karniadakis}}, \bibinfo{journal}{J. Comp. Phys.}
  \textbf{\bibinfo{volume}{227}}, \bibinfo{pages}{2540} (\bibinfo{year}{2008}).

\bibitem[{\citenamefont{Groot and Warren}(1997)}]{Groot_DPD_1997}
\bibinfo{author}{\bibfnamefont{R.~D.} \bibnamefont{Groot}} \bibnamefont{and}
  \bibinfo{author}{\bibfnamefont{P.~B.} \bibnamefont{Warren}},
  \bibinfo{journal}{J. Chem. Phys.} \textbf{\bibinfo{volume}{107}},
  \bibinfo{pages}{4423} (\bibinfo{year}{1997}).

\bibitem[{\citenamefont{Nikunen et~al.}(2003)\citenamefont{Nikunen, Karttunen,
  and Vattulainen}}]{Nikunen_DPD_2003}
\bibinfo{author}{\bibfnamefont{P.}~\bibnamefont{Nikunen}},
  \bibinfo{author}{\bibfnamefont{M.}~\bibnamefont{Karttunen}},
  \bibnamefont{and}
  \bibinfo{author}{\bibfnamefont{I.}~\bibnamefont{Vattulainen}},
  \bibinfo{journal}{Comput. Phys. Commun.} \textbf{\bibinfo{volume}{153}},
  \bibinfo{pages}{407} (\bibinfo{year}{2003}).

\bibitem[{\citenamefont{Thalmanna and Farago}(2007)}]{Thalmanna_TDA_2007}
\bibinfo{author}{\bibfnamefont{F.}~\bibnamefont{Thalmanna}} \bibnamefont{and}
  \bibinfo{author}{\bibfnamefont{J.}~\bibnamefont{Farago}},
  \bibinfo{journal}{J. Chem. Phys.} \textbf{\bibinfo{volume}{127}},
  \bibinfo{pages}{124109} (\bibinfo{year}{2007}).

\bibitem[{\citenamefont{Litvinov et~al.}(2010)\citenamefont{Litvinov, Ellero,
  Hu, and Adams}}]{Litvinov_DPD_2010}
\bibinfo{author}{\bibfnamefont{S.}~\bibnamefont{Litvinov}},
  \bibinfo{author}{\bibfnamefont{M.}~\bibnamefont{Ellero}},
  \bibinfo{author}{\bibfnamefont{X.~Y.} \bibnamefont{Hu}}, \bibnamefont{and}
  \bibinfo{author}{\bibfnamefont{N.~A.} \bibnamefont{Adams}},
  \bibinfo{journal}{J. Comp. Phys.} \textbf{\bibinfo{volume}{229}},
  \bibinfo{pages}{5457} (\bibinfo{year}{2010}).

\bibitem[{\citenamefont{Leimkuhler and Shang}(2015)}]{Leimkuhler_DPD_2015}
\bibinfo{author}{\bibfnamefont{B.}~\bibnamefont{Leimkuhler}} \bibnamefont{and}
  \bibinfo{author}{\bibfnamefont{X.}~\bibnamefont{Shang}}, \bibinfo{journal}{J.
  Comp. Phys.} \textbf{\bibinfo{volume}{280}}, \bibinfo{pages}{72}
  (\bibinfo{year}{2015}).

\bibitem[{\citenamefont{Yamada et~al.}(2018)\citenamefont{Yamada, Itoh,
  Morinishi, and Tamano}}]{Yamada_DPD_2018}
\bibinfo{author}{\bibfnamefont{T.}~\bibnamefont{Yamada}},
  \bibinfo{author}{\bibfnamefont{S.}~\bibnamefont{Itoh}},
  \bibinfo{author}{\bibfnamefont{Y.}~\bibnamefont{Morinishi}},
  \bibnamefont{and} \bibinfo{author}{\bibfnamefont{S.}~\bibnamefont{Tamano}},
  \bibinfo{journal}{J. Chem. Phys.} \textbf{\bibinfo{volume}{148}},
  \bibinfo{pages}{224101} (\bibinfo{year}{2018}).

\bibitem[{\citenamefont{Espa{\~n}ol et~al.}(2016)\citenamefont{Espa{\~n}ol,
  Serrano, Pagonabarraga, and Z{\'u}{\~n}iga}}]{Espanol_ECG_2016}
\bibinfo{author}{\bibfnamefont{P.}~\bibnamefont{Espa{\~n}ol}},
  \bibinfo{author}{\bibfnamefont{M.}~\bibnamefont{Serrano}},
  \bibinfo{author}{\bibfnamefont{I.}~\bibnamefont{Pagonabarraga}},
  \bibnamefont{and}
  \bibinfo{author}{\bibfnamefont{I.}~\bibnamefont{Z{\'u}{\~n}iga}},
  \bibinfo{journal}{Soft Matter} \textbf{\bibinfo{volume}{12}},
  \bibinfo{pages}{4821} (\bibinfo{year}{2016}).

\bibitem[{\citenamefont{L{\'o}pez-Ruiz and Calbet}(2007)}]{Lopez_DMD_2007}
\bibinfo{author}{\bibfnamefont{R.}~\bibnamefont{L{\'o}pez-Ruiz}}
  \bibnamefont{and} \bibinfo{author}{\bibfnamefont{X.}~\bibnamefont{Calbet}},
  \bibinfo{journal}{Am. J. Phys.} \textbf{\bibinfo{volume}{75}},
  \bibinfo{pages}{752} (\bibinfo{year}{2007}).

\bibitem[{\citenamefont{Ripoll}(2002)}]{Ripoll_thesis_2002}
\bibinfo{author}{\bibfnamefont{M.}~\bibnamefont{Ripoll}}, Ph.D. thesis,
  \bibinfo{school}{UNED, Spain} (\bibinfo{year}{2002}).

\bibitem[{\citenamefont{L{\"u}sebrink and Ripoll}(2012)}]{Lusebrink_TIH_2012}
\bibinfo{author}{\bibfnamefont{D.}~\bibnamefont{L{\"u}sebrink}}
  \bibnamefont{and} \bibinfo{author}{\bibfnamefont{M.}~\bibnamefont{Ripoll}},
  \bibinfo{journal}{J. Chem. Phys.} \textbf{\bibinfo{volume}{136}},
  \bibinfo{pages}{084106} (\bibinfo{year}{2012}).

\bibitem[{\citenamefont{Pooley and Yeomans}(2005)}]{Pooley_SRD_2005}
\bibinfo{author}{\bibfnamefont{C.~M.} \bibnamefont{Pooley}} \bibnamefont{and}
  \bibinfo{author}{\bibfnamefont{J.~M.} \bibnamefont{Yeomans}},
  \bibinfo{journal}{J. Phys. Chem. B} \textbf{\bibinfo{volume}{109}},
  \bibinfo{pages}{6505} (\bibinfo{year}{2005}).

\bibitem[{\citenamefont{Yang and Ripoll}(2013)}]{Yang_TPF_2013}
\bibinfo{author}{\bibfnamefont{M.}~\bibnamefont{Yang}} \bibnamefont{and}
  \bibinfo{author}{\bibfnamefont{M.}~\bibnamefont{Ripoll}},
  \bibinfo{journal}{Soft Matter} \textbf{\bibinfo{volume}{9}},
  \bibinfo{pages}{4661} (\bibinfo{year}{2013}).

\bibitem[{\citenamefont{Chapman et~al.}(1990)\citenamefont{Chapman, Cowling,
  and Burnett}}]{Chapman_NUG_1990}
\bibinfo{author}{\bibfnamefont{S.}~\bibnamefont{Chapman}},
  \bibinfo{author}{\bibfnamefont{T.~G.} \bibnamefont{Cowling}},
  \bibnamefont{and} \bibinfo{author}{\bibfnamefont{D.}~\bibnamefont{Burnett}},
  \emph{\bibinfo{title}{The mathematical theory of non-uniform gases}}
  (\bibinfo{publisher}{Cambridge University Press},
  \bibinfo{address}{Cambridge}, \bibinfo{year}{1990}).

\bibitem[{\citenamefont{Guildner}(1975)}]{Guildner_TCG_1975}
\bibinfo{author}{\bibfnamefont{L.~A.} \bibnamefont{Guildner}},
  \bibinfo{journal}{J. Res. Natl. Bur. Stand. A}
  \textbf{\bibinfo{volume}{79A}}, \bibinfo{pages}{407} (\bibinfo{year}{1975}).

\bibitem[{\citenamefont{Hanley et~al.}(1974)\citenamefont{Hanley, McCarty, and
  Haynes}}]{Hanley_VTC_1974}
\bibinfo{author}{\bibfnamefont{H.~J.~M.} \bibnamefont{Hanley}},
  \bibinfo{author}{\bibfnamefont{R.~D.} \bibnamefont{McCarty}},
  \bibnamefont{and} \bibinfo{author}{\bibfnamefont{W.~M.}
  \bibnamefont{Haynes}}, \bibinfo{journal}{J. Phys. Chem. Ref. Data}
  \textbf{\bibinfo{volume}{3}}, \bibinfo{pages}{979} (\bibinfo{year}{1974}).

\bibitem[{\citenamefont{Ziebland and Burton}(1955)}]{Ziebland_TCL_1955}
\bibinfo{author}{\bibfnamefont{H.}~\bibnamefont{Ziebland}} \bibnamefont{and}
  \bibinfo{author}{\bibfnamefont{J.~T.~A.} \bibnamefont{Burton}},
  \bibinfo{journal}{Brit. J. Appl. Phys.} \textbf{\bibinfo{volume}{6}},
  \bibinfo{pages}{416} (\bibinfo{year}{1955}).

\bibitem[{\citenamefont{Noguchi and Gompper}(2008)}]{Noguchi_TCT_2008}
\bibinfo{author}{\bibfnamefont{H.}~\bibnamefont{Noguchi}} \bibnamefont{and}
  \bibinfo{author}{\bibfnamefont{G.}~\bibnamefont{Gompper}},
  \bibinfo{journal}{Phys. Rev. E} \textbf{\bibinfo{volume}{78}},
  \bibinfo{pages}{016706} (\bibinfo{year}{2008}).

\bibitem[{\citenamefont{Lide}(2004)}]{Lide_HBC_2004}
\bibinfo{author}{\bibfnamefont{D.~R.} \bibnamefont{Lide}},
  \emph{\bibinfo{title}{CRC handbook of chemistry and physics}}
  (\bibinfo{publisher}{CRC Press}, \bibinfo{address}{Boca Raton, FL},
  \bibinfo{year}{2004}).

\bibitem[{\citenamefont{Kestin et~al.}(1984)\citenamefont{Kestin, Knierim,
  Mason, Najafi, Ro, and Waldman}}]{Kestin_ETP_1984}
\bibinfo{author}{\bibfnamefont{J.}~\bibnamefont{Kestin}},
  \bibinfo{author}{\bibfnamefont{K.}~\bibnamefont{Knierim}},
  \bibinfo{author}{\bibfnamefont{E.~A.} \bibnamefont{Mason}},
  \bibinfo{author}{\bibfnamefont{B.}~\bibnamefont{Najafi}},
  \bibinfo{author}{\bibfnamefont{S.~T.} \bibnamefont{Ro}}, \bibnamefont{and}
  \bibinfo{author}{\bibfnamefont{M.}~\bibnamefont{Waldman}},
  \bibinfo{journal}{J. Phys. Chem. Ref. Data} \textbf{\bibinfo{volume}{13}},
  \bibinfo{pages}{229} (\bibinfo{year}{1984}).

\bibitem[{\citenamefont{Allen and Tildesley}(1991)}]{Allen_CSL_1991}
\bibinfo{author}{\bibfnamefont{M.~P.} \bibnamefont{Allen}} \bibnamefont{and}
  \bibinfo{author}{\bibfnamefont{D.~J.} \bibnamefont{Tildesley}},
  \emph{\bibinfo{title}{Computer simulation of liquids}}
  (\bibinfo{publisher}{Clarendon Press}, \bibinfo{address}{New York},
  \bibinfo{year}{1991}).

\bibitem[{\citenamefont{{J\"ulich Supercomputing Centre}}(2018)}]{jureca}
\bibinfo{author}{\bibnamefont{{J\"ulich Supercomputing Centre}}},
  \bibinfo{journal}{J. Large-Scale Res. Facil.} \textbf{\bibinfo{volume}{4}},
  \bibinfo{pages}{A132} (\bibinfo{year}{2018}).

\end{thebibliography}

\end{document}